\documentclass[onecolumn]{aastex631}

\usepackage{graphicx}	% Including figure files
\usepackage{amsmath}	% Advanced maths commands
\usepackage{amssymb}	% Extra maths symbols
\usepackage{xcolor}
\usepackage{soul}
\usepackage{float}
\usepackage{lmodern}
\usepackage{placeins}  % 在导言区

\graphicspath{{./}{figures/}}
\shorttitle{AASTeX v6.3.1 Sample article}
\shortauthors{Wang et al.}

\begin{document}

\title{Revisiting the 150 MHz Radio Luminosity Function of Star-Forming Galaxies with LOFAR Deep Fields through a Refined Statistical Framework}

\author[0009-0005-1617-2442]{Wenjie Wang}
\affiliation{Department of Physics, School of Physics and Electronics, Hunan Normal University, Changsha 410081, China}
\affiliation{Key Laboratory of Low Dimensional Quantum Structures and Quantum Control, Hunan Normal University, Changsha 410081, China}
\affiliation{Hunan Research Center of the Basic Discipline for Quantum Effects and Quantum Technologies, Hunan Normal University, Changsha 410081, China}

\author[0000-0001-6861-0022]{Zunli Yuan}
\affiliation{Department of Physics, School of Physics and Electronics, Hunan Normal University, Changsha 410081, China}
\affiliation{Key Laboratory of Low Dimensional Quantum Structures and Quantum Control, Hunan Normal University, Changsha 410081, China}
\affiliation{Hunan Research Center of the Basic Discipline for Quantum Effects and Quantum Technologies, Hunan Normal University, Changsha 410081, China}

\author{Hongwei Yu}
\affiliation{Department of Physics, School of Physics and Electronics, Hunan Normal University, Changsha 410081, China}
\affiliation{Key Laboratory of Low Dimensional Quantum Structures and Quantum Control, Hunan Normal University, Changsha 410081, China}
\affiliation{Hunan Research Center of the Basic Discipline for Quantum Effects and Quantum Technologies, Hunan Normal University, Changsha 410081, China}
\affiliation{Institute of Interdisciplinary Studies, Hunan Normal University, Changsha, Hunan 410081, China}

%\affiliation{Department of Physics and Synergetic Innovation Center for Quantum Effects and Applications, Hunan Normal University, Changsha, Hunan 410081, China}
%\affiliation{Institute of Interdisciplinary Studies, Hunan Normal University, Changsha, Hunan 410081, China}

\author[0000-0003-2721-2559]{Yang Liu}
\affiliation{Purple Mountain Observatory, Chinese Academy of Sciences, Nanjing 210023, China}

\author[0000-0003-2341-9755]{Yu Luo}
\affiliation{Department of Physics, School of Physics and Electronics, Hunan Normal University, Changsha 410081, China}
\affiliation{Key Laboratory of Low Dimensional Quantum Structures and Quantum Control, Hunan Normal University, Changsha 410081, China}
\affiliation{Hunan Research Center of the Basic Discipline for Quantum Effects and Quantum Technologies, Hunan Normal University, Changsha 410081, China}

\author{Puxun Wu}
\affiliation{Department of Physics, School of Physics and Electronics, Hunan Normal University, Changsha 410081, China}
\affiliation{Key Laboratory of Low Dimensional Quantum Structures and Quantum Control, Hunan Normal University, Changsha 410081, China}
\affiliation{Hunan Research Center of the Basic Discipline for Quantum Effects and Quantum Technologies, Hunan Normal University, Changsha 410081, China}

\correspondingauthor{Zunli Yuan}
\email{yzl@hunnu.edu.cn}
\correspondingauthor{Hongwei Yu}
\email{hwyu@hunnu.edu.cn}

\begin{abstract}
We present a comprehensive analysis of the 150~MHz radio luminosity function (LF) of star-forming galaxies (SFGs) using deep observations from the LOFAR Two-metre Sky Survey  in the ELAIS-N1, Bo\"{o}tes, and Lockman Hole fields. Our sample comprises $\sim$56,000 SFGs over $0 < z < 5.7$. We first analyze the deepest field (ELAIS-N1), then jointly model all three fields while accounting for their distinct flux limits and selection functions. Using adaptive kernel density estimation (KDE), we reconstruct the LF continuously across redshift and luminosity without  binning or parametric assumptions. The KDE results reveal clear signatures of joint luminosity and density evolution (LADE). Motivated by this, we construct and fit three parametric models—pure luminosity evolution (PLE) and two LADE variants—using a full maximum-likelihood method that includes completeness corrections and constraints from the local radio LF and Euclidean-normalized source counts (SCs). Model selection using Akaike and Bayesian Information Criteria strongly favors LADE over PLE. For ELAIS-N1, the more flexible LADE model (Model C) provides the best fit, while for the combined fields, the simpler Model B balances fit quality and complexity more effectively. Both LADE models reproduce the observed LFs and SCs across luminosity and flux density ranges, whereas PLE underperforms. We also identify a mild excess at the bright end of the LF, likely due to residual AGN contamination. This study demonstrates that combining KDE with parametric modeling offers a robust framework for quantifying the evolving radio LF of SFGs, paving the way for future work with next-generation surveys like the SKA.
\end{abstract}


\section{Introduction}
\label{sec_intro}

Star formation is one of the most fundamental processes driving galaxy evolution across cosmic time. Understanding when and how galaxies assemble their stellar mass offers key insight into the mechanisms that shape their histories, from the epoch of reionization ($z\gtrsim6$) through the peak of cosmic activity at $z\sim2$ and into the low‐redshift Universe.

The star formation rate (SFR) can be traced across the electromagnetic spectrum \citep[e.g.,][]{kennicutt1998star}.   Ultraviolet (UV) observations directly measure the light from newly formed massive stars and thus provide a sensitive probe of star formation out to high redshifts \citep[$z \sim 11$; e.g.,][]{Mclure_2013, Oesch_2018, Bouwens_2021}. However, UV-based SFR estimates require substantial corrections for dust extinction  \citep{Smail_1997,Riechers_2013}. The absorbed UV photons are  re-emitted at far-infrared (FIR) and submillimeter wavelengths, providing a complementary view of obscured star formation \citep{kennicutt1998star}. Unfortunately,  FIR and submillimeter surveys often suffer from low angular resolution, complicating the identification of faint  sources.  In recent years, additional tracers of star formation have emerged, including optical emission lines and radio continuum surveys \citep{Ouchi-2010,Drake-2013,Schober-2015,Aird-2017}.

Continuum radio observations offer a dust-unbiased view of ongoing star formation activity. Short-lived massive stars end their lives as supernovae whose remnants accelerate cosmic-ray electrons, producing synchrotron radiation detectable at frequencies  below 30~GHz \citep[e.g.,][]{Sadler_1989,Tabatabaei_2017}. This emission correlates strongly with the infrared luminosity from star-forming regions, giving rise to the well-established far-infrared–radio correlation \citep[e.g.,][]{Condon-1991,Bell_2003}, which holds over a broad range of galaxy luminosities and redshifts. Radio emission thus provides a powerful, extinction-free tracer of the SFR in galaxies \citep{magnelli2015far, Calistro_2017, Delvecchio_2017, Algera_2020}.

Across the full suite of multiwavelength surveys described above, a critical quantity that can be derived is the cosmic star formation rate density (SFRD), defined as the total SFR per unit comoving volume. Observations show that the SFRD rises from high redshift, peaks around $z \sim 2$, and declines by an order of magnitude toward the present day \citep{Madau_1996MNRAS.283.1388M,Haarsma_2000,Madau_2014,2017A&A...602A...5N,2022ApJ...941...10V,2023MNRAS.523.6082C,2024A&A...683A.174W}. Nevertheless, the precise behavior of the SFRD beyond $z \sim 3$ remains uncertain. UV-based studies suggest a steep decline \citep{2015ApJ...803...34B,2016MNRAS.459.3812M,2018ApJ...854...73I}, whereas radio and submillimeter surveys sometimes indicate a more gradual decrease \citep{Gruppioni_2013,2020A&A...643A...8G,RowanRobinson_2016,2017A&A...602A...5N,2022ApJ...927..204E}. This tension points to the need for complementary, multiwavelength approaches to mitigate observational biases and better understand the underlying physics.

Traditional radio surveys at 1.4~GHz can already probe star formation over a wide range of redshift. However, since radio emission is redshifted to lower frequencies at high redshift, observations at 150~MHz provide a complementary view of star formation in the early Universe \citep{2010Natur.468..772P}. Consequently, several recent efforts have shifted toward low-frequency radio surveys to assess the universal validity of radio-based SFR indicators \citep[e.g.,][]{2018MNRAS.475.3010G, 2020MNRAS.491.5911O, 2021A&A...648A...6S}. Deep 150~MHz observations from the Low Frequency Array (LOFAR) Two-metre Sky Survey (LoTSS) now routinely reach $\mu$Jy sensitivities, enabling statistically robust measurements of star-forming galaxies (SFGs) out to $z\gtrsim5$ \citep[e.g.,][]{2023MNRAS.523.6082C}.

To quantify the cosmic star formation history in a physically meaningful way, one requires accurate measurements of the SFRD across redshift. A foundational ingredient for such measurements is the radio luminosity function (LF) of SFGs, which describes the number density of galaxies as a function of luminosity and redshift.
By measuring the LF and integrating it over the full luminosity range, one can derive the SFRD at each epoch. Early studies often assumed pure luminosity evolution (PLE) \citep[e.g.,][]{2009ApJ...690..610S,2017A&A...602A...5N,2020MNRAS.491.5911O,2022MNRAS.509.4291M}, whereas more recent analyses have considered models combining both luminosity and density evolution (LADE) \citep[][]{2022ApJ...941...10V,2023MNRAS.523.6082C, 2024A&A...683A.174W}. Most of these works estimate the LF using the classical $1/V_{\mathrm{max}}$ method \citep{1968ApJ...151..393S}, which provides a convenient but binned representation of the underlying distribution.

In this work, we revisit the 150~MHz radio LF of SFGs using the same deep LOFAR observations from the ELAIS-N1, Bo\"{o}tes, and Lockman Hole fields as in \citet{2023MNRAS.523.6082C}. While that study represents a major step forward in assembling large, well-characterized SFG samples at low radio frequencies, it relies on a two-step procedure: deriving LF points using the $1/V_{\mathrm{max}}$ method, and subsequently fitting those binned values with parametric models. This indirect approach, though widely used \citep[e.g.,][]{smolvcic2009cosmic, 2017A&A...602A...5N, ocran2020cosmic, 2022MNRAS.509.4291M, 2022ApJ...941...10V}, introduces binning-related uncertainties and does not make full use of the statistical information encoded in the unbinned data \citep{fan2001high}.

To overcome these limitations, we adopt a statistically rigorous framework that combines non-parametric and parametric approaches. Specifically, we first apply adaptive kernel density estimation (KDE) \citep{yuan2020flexible,yuan2022flexible} to reconstruct the LF continuously across redshift and luminosity. Guided by the empirical trends revealed by the KDE, we then construct physically motivated LF models and constrain them using a full maximum-likelihood method that incorporates completeness corrections and additional constraints from the local radio LF and Euclidean-normalized source counts (SCs). This approach enables a more accurate and robust characterization of the evolving radio LF and maximizes the scientific return from existing deep surveys.

This paper is structured as follows. Section~\ref{sec_sample} describes the LOFAR data and multiwavelength ancillary catalogues. Section~\ref{sec_methods} outlines our KDE framework and maximum-likelihood modeling approach. The resulting LFs and their evolution are presented in Section~\ref{sec_results}, while Section~\ref{sec_discussion} discusses the implications of our findings and summarizes our conclusions. Throughout, we adopt a flat $\Lambda$CDM cosmology with $H_0 = 70~\mathrm{km\,s^{-1}\,Mpc^{-1}}$, $\Omega_\Lambda = 0.7$, and $\Omega_m = 0.3$, and assume a power‐law radio spectrum $F_\nu \propto \nu^{-\alpha}$ with $\alpha = 0.7$ where required.

\section{Sample}
\label{sec_sample}

\begin{figure*}
	\centering
	\includegraphics[width=0.33\textwidth]{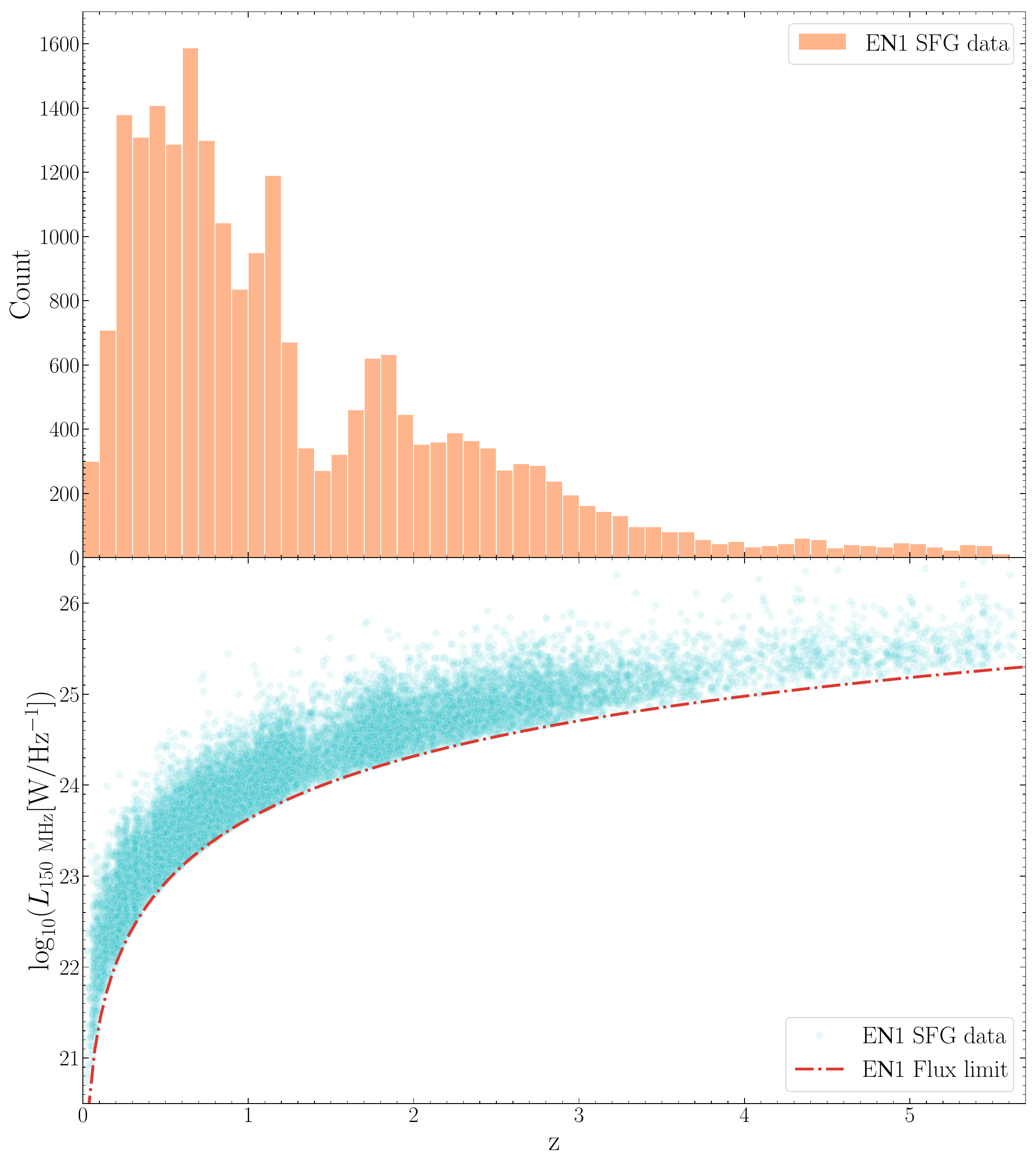}
	\hspace{-0.1in}
	\includegraphics[width=0.33\textwidth]{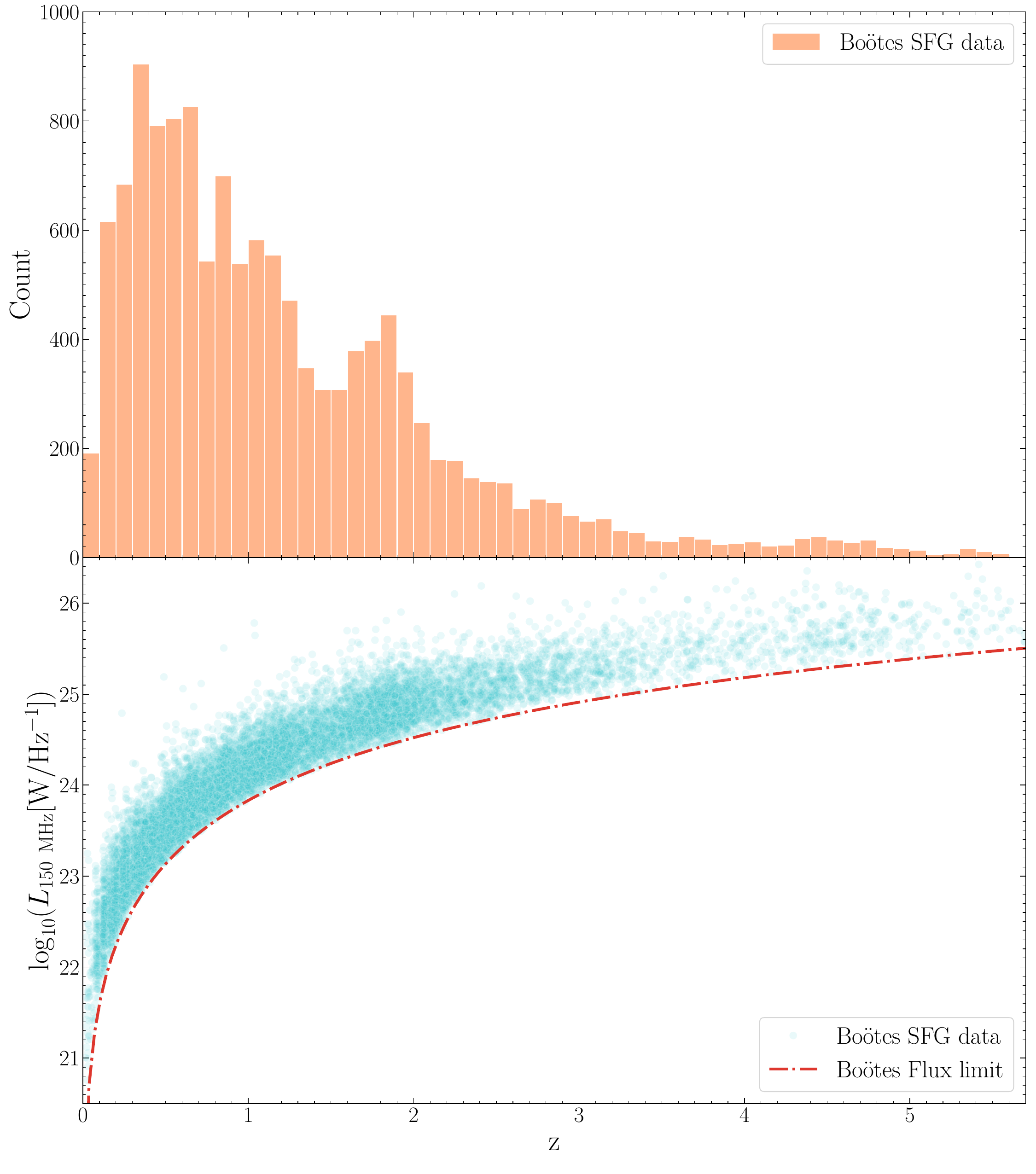}
    \hspace{-0.1in}
	\includegraphics[width=0.33\textwidth]{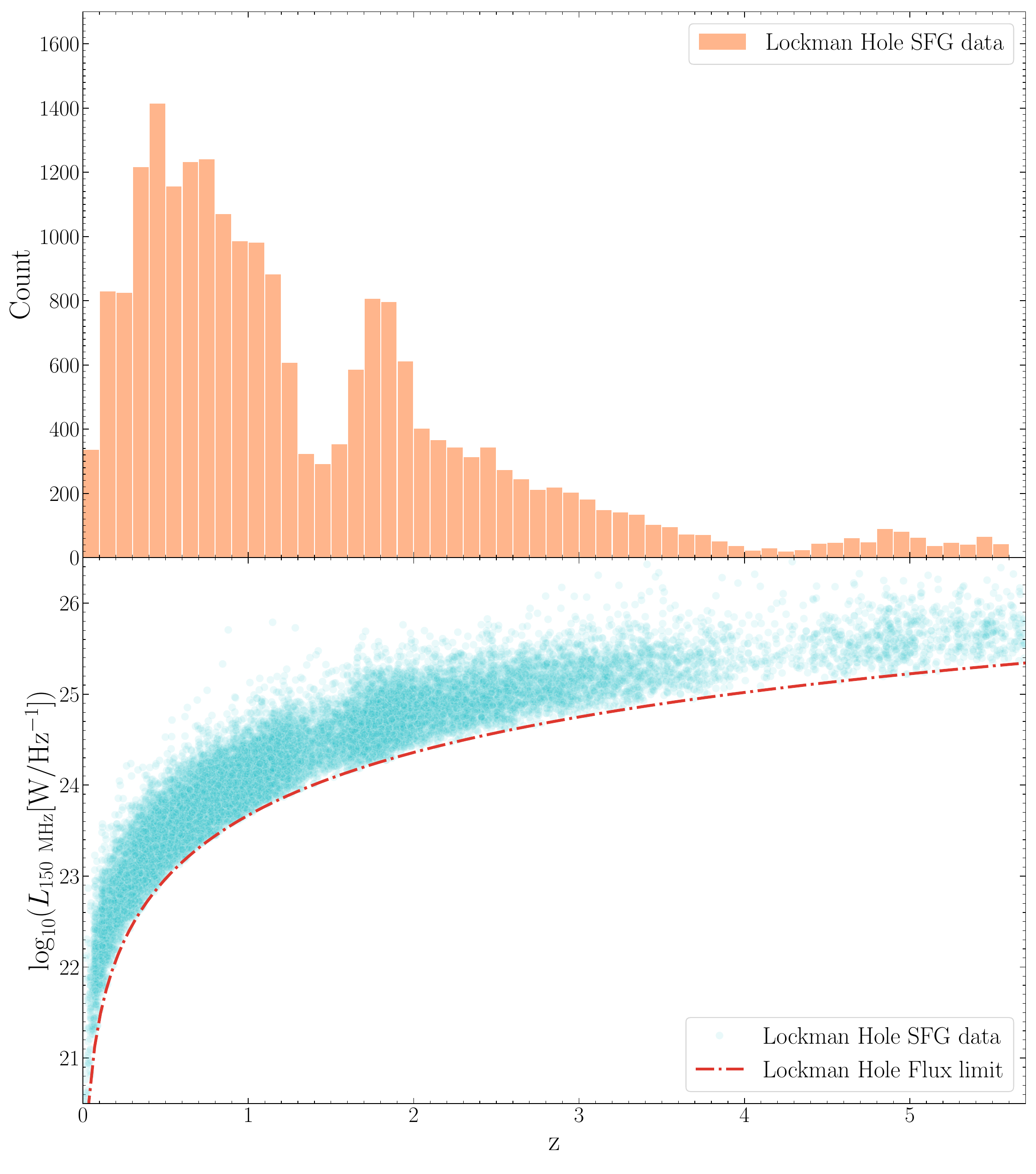}
	\caption{
    Redshift distribution ($top$) and scatter plot ($bottom$) of our SFG sample for three fields: ELAIS-N1 ($left$), Bo{\"o}tes ($middle$), and Lockman Hole ($right$). The red dashed lines indicate the flux limits of $F_{150\,\rm{MHz}}=100~\mu\rm{Jy}$ for ELAIS-N1, $160~\mu\rm{Jy}$ for Bo{\"o}tes, and $110~\mu\rm{Jy}$ for Lockman Hole. Note that a small number of sources fall below these flux limit lines and are excluded from our subsequent analysis.
    }
	\label{fig:sample_data}
\end{figure*}

In this work, we use the same sample of SFGs presented in \cite{2023MNRAS.523.6082C}, which was derived from deep $150\,\rm{MHz}$ observations of the ELAIS-N1, Bo{\"o}tes, and Lockman Hole fields, conducted as part of the LOFAR Two-metre Sky Survey (LoTSS) Deep Fields project.
The data analysis and multi-band association procedure are fully described in \cite{2023MNRAS.523.6082C}, and readers are referred to this work for further details. Here we summarize some key points about the sample.

The radio data were obtained using the LOFAR High-Band Antenna (HBA) array, with integration times of $164\,\rm{hr}$, $80\,\rm{hr}$, and $112\,\rm{hr}$ for ELAIS-N1, Bo{\"o}tes, and Lockman Hole, respectively, achieving rms sensitivities of $20\,\mu\rm{Jy/beam}$, $32\,\mu\rm{Jy/beam}$, and $22\,\mu\rm{Jy/beam}$ at the pointing centers \citep{Tasse2021,Sabater2021}. The observations cover areas of approximately $25\,\rm{deg}^{2}$ per field (within the 30\% power point), with an angular resolution of $6''$. Source extraction was performed using PyBDSF \citep{Mohan2015}, yielding initial catalogs of $84,862$, $36,767$, and $50,112$ radio sources for ELAIS-N1, Bo{\"o}tes, and Lockman Hole, respectively.

This sample is complemented by extensive multiwavelength data from UV to FIR, compiled and cross-matched by \cite{Kondapally2021}.
Photometric coverage includes UV data from GALEX \citep{Martin2005}, optical data from surveys such as Pan-STARRS \citep{Chambers2016} and NDWFS \citep{1999ASPC..191.....W}, near-IR data from UKIDSS-DXS \citep{Lawrence2007} and NEWFIRM \citep{Gonzalez2010}, and mid-to-far-IR data from ${\it{Spitzer}}$ \citep{Lonsdale2003,Ashby2009} and ${\it{Herschel}}$ \citep{Oliver2012}.
Photometric redshifts were derived using a hybrid approach combining template fitting and machine learning, as detailed in \cite{Duncan2021}, with spectroscopic redshifts available for $\sim8.6\%$ of sources.
Cross-matching of radio sources to multiwavelength counterparts was limited to regions with optimal photometric coverage: $6.74\,\rm{deg}^{2}$ in ELAIS-N1, $8.63\,\rm{deg}^{2}$ in Bo{\"o}tes, and $10.28\,\rm{deg}^{2}$ in Lockman Hole \citep{Kondapally2021}.
Classification of SFGs and active galactic nuclei (AGN) was performed using multiple methods, including SED fitting with codes such as {\small{MAGPHYS}}, {\small{BAGPIPES}}, {\small{AGNfitter}}, and {\small{CIGALE}} \citep{Best2023}, alongside radio-excess criteria based on the $150\,\rm{MHz}$ luminosity-to-SFR relation. SFGs were defined as sources lacking significant AGN signatures in optical/IR/X-ray and radio excess. For a detailed account of the observations, data processing, and sample selection, see  \cite{2023MNRAS.523.6082C}, \cite{Tasse2021}, \cite{Sabater2021}, \cite{Kondapally2021}, and \cite{Best2023}.

After cross-matching and filtering, the final sample comprises $55,991$ SFGs with redshifts ranging from $0.001 < z < 5.7$, including $21,725$ in the ELAIS-N1 field, $12,895$ in the Bo{\"o}tes field, and $21,371$ in the Lockman Hole field. All the sources have (spectroscopic or photometric) redshifts and the rest-frame 150 MHz luminosities. The redshift distribution of the SFG sample as well as its scatter plot are shown in Figure \ref{fig:sample_data}. 
The red dashed curve indicates the 150 MHz flux limit line defined as
\begin{eqnarray}
        \label{flim}
        f_{\rm{lim ~ 150 MHz}}(z)=\frac{4 \pi D_{L}^{2}}{(1+z)^{1-\alpha}}F_{\rm{lim ~ 150 MHz}},
\end{eqnarray}
where $D_L$ represents the luminosity distance at redshift $z$, and $\alpha=0.7$ is adopted for the K-correction. For illustration, we take the nominal central sensitivities of the LoTSS Deep Fields as $F_{\rm{lim ~ 150 MHz}} = 100~\mu\rm{Jy}$ for ELAIS-N1, $160~\mu\rm{Jy}$ for Bo{\"o}tes, and $110~\mu\rm{Jy}$ for Lockman Hole (Cochrane et al. 2023). Since the noise increases away from the pointing centre, these values do not represent the true local $5\sigma$ threshold across the entire field. Instead, we use them as conservative reference limits to exclude sources lying below the corresponding flux–redshift curve, so that our analysis focuses on the most complete part of the survey.

\section{Methods}
\label{sec_methods}
The LF, \( \Phi(z, L) \), is defined as the number of sources per unit comoving volume \( V(z) \) per unit logarithmic luminosity interval \( \log L \), i.e.,
\begin{eqnarray}
\label{eq:LFdf}
\Phi(z,L) = \frac{d^{2}N}{dV\,d\log L}.
\end{eqnarray}

Methods for estimating the LF can be broadly classified into two categories: parametric and non-parametric approaches. Non-parametric methods, such as the classical \(1/V_{\max}\) estimator \citep{1968ApJ...151..393S} and the more recent KDE method proposed by \citet{yuan2020flexible, yuan2022flexible}, are data-driven and make minimal assumptions about the underlying form of the LF. They are particularly useful for revealing unexpected structures in the data without being constrained by pre-imposed model forms.

Parametric methods, by contrast, assume a specific functional form for the LF and typically rely on maximum-likelihood estimation (MLE) to determine model parameters. When the assumed model closely resembles the true underlying distribution, this approach can yield highly accurate and efficient estimates. However, in practice, we seldom know the true form of the LF a priori, and inappropriate model choices may lead to biased results.

Therefore, it is often advantageous to combine both approaches. We can first use a non-parametric method (e.g., KDE) to obtain a smooth, empirical estimate of the LF shape, and then use this insight to guide the selection of a suitable parametric model. In the following, we briefly introduce the non-parametric KDE method and the parametric maximum-likelihood approach adopted in this work.

\subsection{KDE-Based Luminosity Function Estimation}

Mathematically, KDE is a statistically rigorous, nonparametric approach for reconstructing continuous density functions without assuming a specific parametric form, and has been extensively studied in the statistical literature \citep[e.g.][]{wasserman2006all,chen2017tutorial,davies2018fast}. Based on the principles of KDE, \citet{yuan2020flexible, yuan2022flexible} proposed a flexible framework for estimating LFs, hereafter referred to as the KDE method. This method does not require any model assumptions, since it generates the LF relying only on the available data.

Given a two-dimensional dataset of redshift and luminosity points, $(z_i, L_i)$ for $i=1,2,...,n$, directly applying KDE can introduce boundary biases due to truncation. To mitigate this, a transformation-reflection method is employed, mapping the data into an unbounded space:
\begin{eqnarray}
\label{eq:trans}
x = \ln\left(\frac{z-Z_1}{Z_2-z}\right), \quad y = L - f_{\mathrm{lim}}(z),
\end{eqnarray}
where $f_{\mathrm{lim}}(z)$ denotes the luminosity truncation boundary at redshift $z$.

The KDE method is then applied to the transformed data in the \( (x, y) \) space. The density of $(x, y)$, denoted as $\hat{f}(x, y)$, can be estimated by
\begin{eqnarray}
\label{trandrf3}
\begin{aligned}
\hat{f}(x, y) =\frac{1}{n h_{1} h_{2}}
\times \sum_{j=1}^{n}\left(K\left(\frac{x-x_{j}}{h_{1}}, \frac{y-y_{j}}{h_{2}}\right)+K\left(\frac{x-x_{j}}{h_{1}}, \frac{y+y_{j}}{h_{2}}\right)\right)
\end{aligned}
\end{eqnarray}

To further improve the estimation accuracy, especially in cases where the data distribution is highly inhomogeneous, the density of $(x, y)$ can also be estimated using adaptive bandwidths, leading to the so-called adaptive KDE, denoted as $\hat{f}_{\mathrm{a}}(x, y)$. In this approach, the smoothing parameters are allowed to vary across different data points depending on the local density, such that smaller bandwidths are used in dense regions and larger ones in sparse regions. Specifically, a pilot estimate $\tilde{f}(x, y)$ is first obtained using the fixed-bandwidth KDE. Then, the local bandwidths are modulated according to
\begin{eqnarray}
\lambda_1(x_j, y_j) = h_{10} \tilde{f}(x_j, y_j)^{-\beta}, \quad \lambda_2(x_j, y_j) = h_{20} \tilde{f}(x_j, y_j)^{-\beta},
\end{eqnarray}
where $h_{10}$ and $h_{20}$ are global scaling factors and $\beta$ is a sensitivity parameter controlling the degree of adaptation ($0 \leq \beta \leq 1$). This scheme enables the estimator to adaptively balance bias and variance across the domain, effectively mitigating both oversmoothing in dense areas and undersmoothing in sparse regions.

Once the adaptive bandwidths are determined, the density of $(x, y)$ can be estimated by an adaptive KDE, denoted as $\hat{f}_{\mathrm{a}}(x, y)$. Its mathematical form follows the same structure as Equation~(\ref{trandrf3}), with the fixed bandwidths $h_1$ and $h_2$ replaced by the location-dependent adaptive bandwidths $\lambda_1(x_j, y_j)$ and $\lambda_2(x_j, y_j)$ for each kernel term.

Transforming back to the original coordinates, the density of $(z, L)$ can be written as
\begin{eqnarray}
\hat{p}_{\mathrm{a}}(z, L) = \hat{f}_{\mathrm{a}}(x, y) \cdot \frac{Z_2 - Z_1}{(z - Z_1)(Z_2 - z)},
\end{eqnarray}
and the adaptive estimate of the LF becomes
\begin{eqnarray}
\label{adaptiveLF}
\hat{\phi}_{\mathrm{a}}(z,L)=\frac{n(Z_2-Z_1)\hat{f}_{\mathrm{a}}(x,y)}{(z-Z_1)(Z_2-z)\Omega\frac{dV}{dz}},
\end{eqnarray}
where $dV/dz$ is the differential comoving volume per unit solid angle, and $\Omega$ is the survey solid angle. The optimal values of $h_{10}$, $h_{20}$, and $\beta$ are obtained by minimizing a likelihood cross-validation (LCV) \textit{objective function} tailored to the transformation-reflection framework \citep[for the full mathematical treatment, see][]{yuan2022flexible}.

Due to the variation of the noise along the map, the sample's true detection boundary is not described by a single flux limit (e.g., $100~\mu Jy$ or ELAIS-N1). Instead, the actual flux limit varies spatially across the map. As indicated by Equation~(\ref{eq:trans}), our KDE method is currently capable of handling only a single, well-defined flux limit. This simplification inevitably introduces bias near the boundary, as it cannot fully account for the `fuzzy' detection threshold caused by spatially varying noise. To mitigate this boundary bias, we assign each source a weight $w_i = p(z_i, L_i)$ in the KDE computation, where $p(z, L)$ represents the survey selection function, whose explicit form will be given in Equation~(\ref{p_zL}). Accordingly, we adopt the weighted form of the adaptive KDE in our analysis \citep[Equation (C3)][]{yuan2022flexible}.

\subsection{Parametric Luminosity Function Modeling}
\label{methods_LFLF}

While non-parametric methods such as KDE provide flexible, data-driven estimates of the LF, they do not directly yield a parametric form that can be used for extrapolation or theoretical interpretation. To this end, we also adopt a parametric approach based on MLE, which allows for the inference of model parameters within an assumed analytical form of the LF.

Given a model LF $\Phi(z, L\,|\,\boldsymbol{\theta})$ parameterized by $\boldsymbol{\theta}$, the optimal parameter values can be obtained by maximizing the likelihood function constructed from the observed data. Following the formalism of \citet{marshall1983analysis} and \citet{fan2001high}, the negative log-likelihood for a single field is given by
\begin{eqnarray}
\label{eq:likelihood1}
\begin{aligned}
S_{\rm single} = -2 \sum_{i=1}^{n} \ln[\Phi(z_i, L_i)\,p(z_i, L_i)]
+ 2 \int\!\!\!\!\int_{W} \Phi(z, L)\,p(z, L)\,\Omega[F(z)]\,\frac{dV}{dz}\,dz\,dL,
\end{aligned}
\end{eqnarray}
where $p(z, L)$ is the selection probability of a SFG at redshift $z$ and luminosity $L$, and $W$ denotes the effective survey region in the $(z, L)$ plane. The first term evaluates the likelihood at the observed data points, while the second term accounts for the expected number of sources over the survey volume. $\Omega[F(z)]$ is the sky area over which a source with flux $F$ can be detected at 5$\sigma$ significance \citep{Kondapally2021}, and $dV/dz$ is the differential comoving volume per unit solid angle \citep{hogg1999distance}.

For our SFG sample, the selection function $p(z,L)$ is modeled as
\begin{equation}
\label{p_zL}
p(z,L)= C_{\text{radio}}[F(z)] \times C_{\text{photometric}}(z),
\end{equation}
where $C_{\text{radio}}[F(z)]$ describes the radio completeness of the LOFAR catalogue as a function of flux density, and $C_{\text{photometric}}(z)$ is an empirical correction for photometric redshift errors such as aliasing \citep{2023MNRAS.523.6082C}. We adopt the tabulated values and procedures from \citet{2023MNRAS.523.6082C}, and refer the reader to their Fig. 2 and Fig. 3 for details.

Equation~(\ref{eq:likelihood1}) applies to a single field with well-defined completeness. In this work, we apply it specifically to the ELAIS-N1 field, which benefits from the deepest radio observations among the three. For a more comprehensive analysis, we also perform a joint fit combining all three sub-samples—ELAIS-N1, Bo{\"o}tes, and Lockman Hole. Given the differences in flux limits, survey areas, and selection functions, these fields cannot be merged at the catalogue level. Instead, their contributions are combined at the likelihood level, with each field treated using its respective completeness and sky coverage. The resulting total negative log-likelihood is
\begin{eqnarray}
\label{eq:likelihood2}
\begin{aligned}
S_{\rm all} = -2 \sum_{j} \sum_{i=1}^{n_j} \ln[\Phi_j(z_i, L_i)\,p_j(z_i, L_i)]
+ 2 \sum_{j} \int\!\!\!\!\int_{W_j} \Phi_j(z, L)\,p_j(z, L)\,\Omega_j[F(z)]\,\frac{dV}{dz}\,dz\,dL,
\end{aligned}
\end{eqnarray}
where the index $j$ runs over the three fields, $n_j$ is the number of sources in field $j$, and all quantities—$\Phi_j$, $p_j$, $\Omega_j$, $W_j$—are field dependent. In our implementation, we consider both field-dependent LFs ($\Phi_j$ varying with $j$), and a simplified scenario where all three fields share a common LF, $\Phi_j \equiv \Phi$, to improve statistical constraints while accounting for field-specific incompleteness.

For the selection function $p(z, L)$ in both Equations~(\ref{eq:likelihood1}) and (\ref{eq:likelihood2}), we follow the interpolation-based approach developed in our previous work \citep{2024A&A...683A.174W}, which effectively captures the complex selection boundaries in the $(z, L)$ plane.

While Equations~(\ref{eq:likelihood1}) and (\ref{eq:likelihood2}) provide a statistical framework for fitting the $(z, L)$ distribution of the sample, they are not the final objective functions adopted in this study. 
To further improve the parameter constraints and break degeneracies in redshift and luminosity, we incorporate two additional one-dimensional observational constraints: (1) the local radio luminosity function (LRLF) of SFGs, and (2) the Euclidean-normalized radio source counts (SCs) in the ELAIS-N1 field. 
This choice is motivated by the limited depth of our radio data: even in the deepest ELAIS-N1 field, the sample does not reach the knee of the LF at $z\gtrsim0.5$, leaving the faint-side normalization and slope poorly constrained when relying on the catalog likelihood alone. 
In such an incompletely sampled regime, strong degeneracies arise between $\phi_\star$, $L_\star$, and the redshift-evolution parameters. 
The independently measured LRLF at $z\simeq0.1$ provides a robust local anchor for the low-luminosity end, while the Euclidean-normalized SCs add complementary information by constraining the luminosity- and redshift-integrated abundance and are particularly sensitive to the space density of faint systems. 
Assuming Gaussian measurement uncertainties, including these data as additional $\chi^2$ terms is equivalent to multiplying the corresponding likelihoods, thereby regularizing the fit with well-motivated external information and preventing unphysical extrapolation.
Following \citet{willott2001radio} and \citet{yuan2017mixture}, we implement these constraints as additional $\chi^2$ penalty terms in the likelihood function, each defined as
\begin{eqnarray}
\label{eq:chi2}
\chi^{2} = \sum_{i=1}^{n} \left( \frac{f_{\mathrm{data},i} - f_{\mathrm{mod},i}}{\sigma_{\mathrm{data},i}} \right)^2,
\end{eqnarray}
where $f_{\mathrm{data},i}$ is the observed value in the $i$th bin, $f_{\mathrm{mod},i}$ is the corresponding model prediction, and $\sigma_{\mathrm{data},i}$ is the observational uncertainty.
For the SCs, the term $f_{\mathrm{mod},i}$ is obtained from Equation~(\ref{eq:sc1}) by integrating the evolving LF over luminosity and redshift, while for the LRLF it is given by Equation~(\ref{eq:LF2}), corresponding to the Saunders et al. (1990) form evaluated at $z=0.10$.

The final objective functions used for parameter estimation are given by:
\begin{eqnarray}
\label{eq:chi2_2}
\begin{aligned}
S_{\text{single final}} &= S_{\text{single}} + \chi^2_{\text{LRLF}} + \chi^2_{\text{SC EN1}}, \\
S_{\text{all final}} &= S_{\text{all}} + \chi^2_{\text{LRLF}} + \chi^2_{\text{SC all}},
\end{aligned}
\end{eqnarray}
where $\chi^2_{\text{LRLF}}$ and $\chi^2_{\text{SC}}$ denote the $\chi^2$ contributions from the LRLF and SCs, respectively. These augmented likelihood functions serve as the final basis for model fitting in both the single-field and combined-field analyses.

To evaluate $\chi^2_{\text{LRLF}}$, we use the binned measurements of the LRLF at 150\,MHz from \citet{2023MNRAS.523.6082C}, based on deep LOFAR observations at $z \lesssim 0.1$. The data points used in our modeling are shown in Figure~\ref{fig:LRLF}.

To compute $\chi^2_{\text{SC EN1}}$ and $\chi^2_{\text{SC all}}$, we derive the 150\,MHz Euclidean-normalized SCs using flux density distributions from the ELAIS-N1 field and the combined sample across all fields, respectively, both based on \citet{2023MNRAS.523.6082C}. These binned measurements help constrain the global normalization and redshift evolution of the LF. The resulting SCs are shown in Figure~\ref{fig:sc}.

Following \citet{Padovani_2016} and \citet{yuan2017mixture}, the differential SCs $n(F_\nu)$ are related to the evolving LF $\Phi(z, L)$ via
\begin{eqnarray}
\label{eq:sc1}
\frac{n(F_\nu)}{4\pi} = 4\pi\,\frac{c}{H_0} \int_{z_\text{min}(F_\nu)}^{z_\text{max}(F_\nu)}
\frac{\Phi(z, L(F_\nu, z))\,D_L^4(z)}{(1+z)^{3 - \alpha} \sqrt{\Omega_\mathrm{m}(1+z)^3 + \Omega_\Lambda}}\,dz,
\end{eqnarray}
where $c$ is the speed of light, $D_L(z)$ is the luminosity distance, $\alpha$ is the spectral index, and $z_\text{min}$ and $z_\text{max}$ denote the redshift integration limits for a given flux density $F_\nu$.

Using Equation~(\ref{eq:chi2_2}), we determine the best-fit LF parameters by minimizing $S_{\text{single final}}$ and $S_{\text{all final}}$, respectively. The parameter space is explored using a Markov Chain Monte Carlo (MCMC) method with uniform (i.e., uninformative) priors, following the Bayesian framework of \citet{yuan2016mixture}, and implemented via the {\sc emcee} sampler \citep{foreman2013emcee}.

\begin{figure}
	\centering
	\includegraphics[width=0.7\columnwidth]{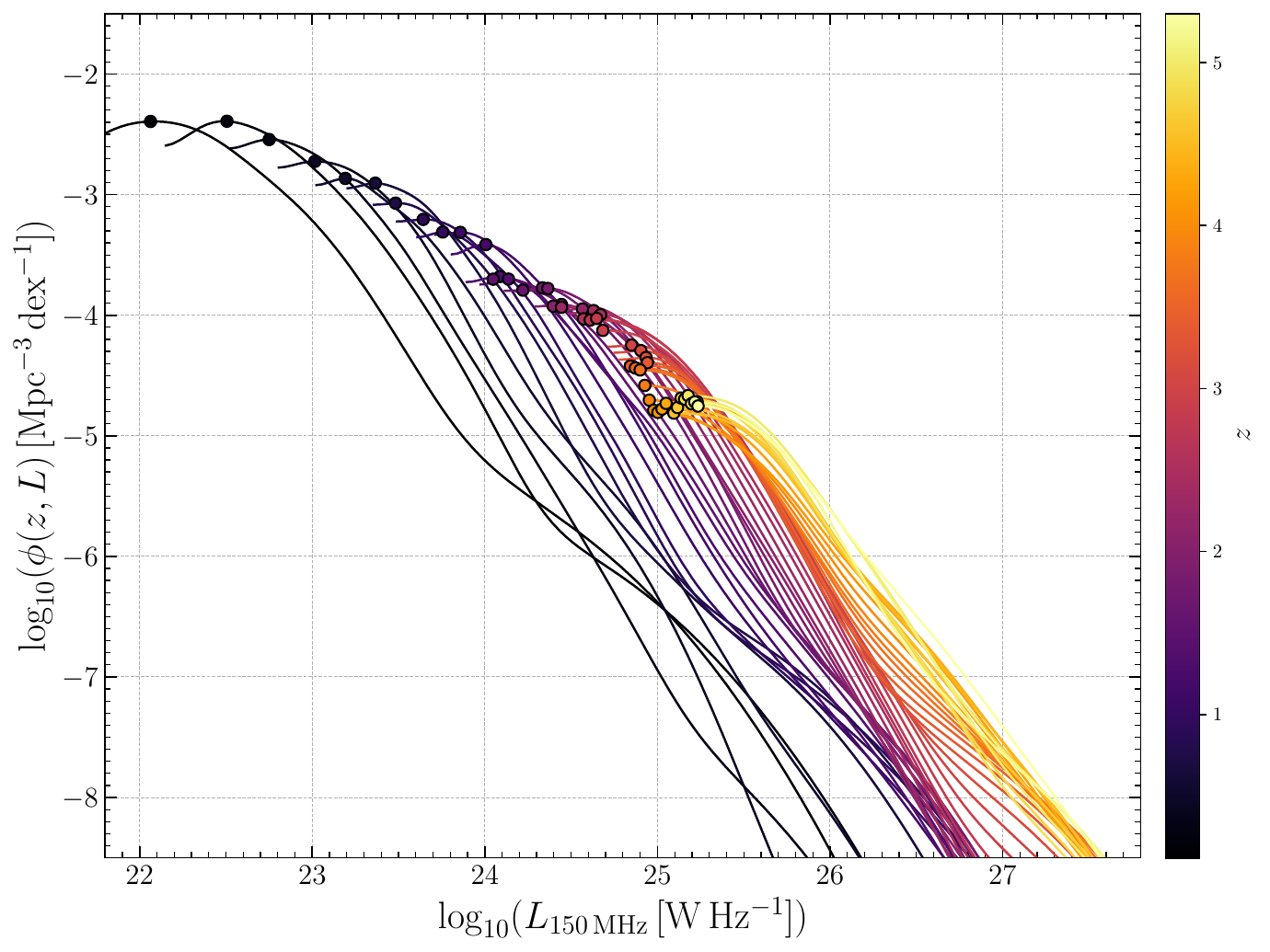}
    \caption{LFs estimated at a series of redshift grid points using the adaptive KDE method. The resulting curves are color-coded according to redshift. Solid circles indicate the flattest regions of the LF at each redshift.}
	\label{fig:KDE_DELE}
\end{figure}

\begin{figure*}
        \centering
        \includegraphics[width=0.5\textwidth]{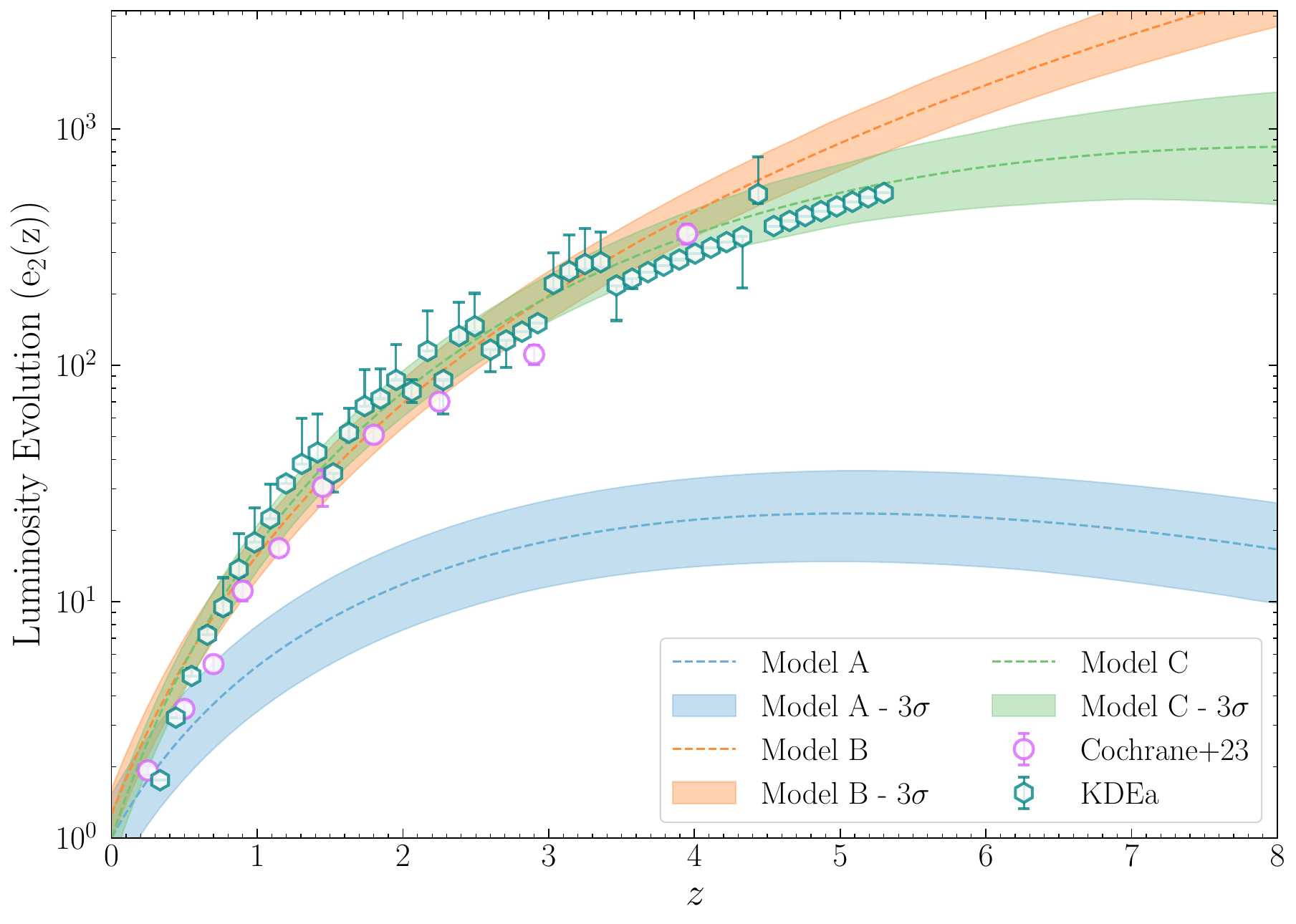}
        \hspace{-0.1in}
        \includegraphics[width=0.5\textwidth]{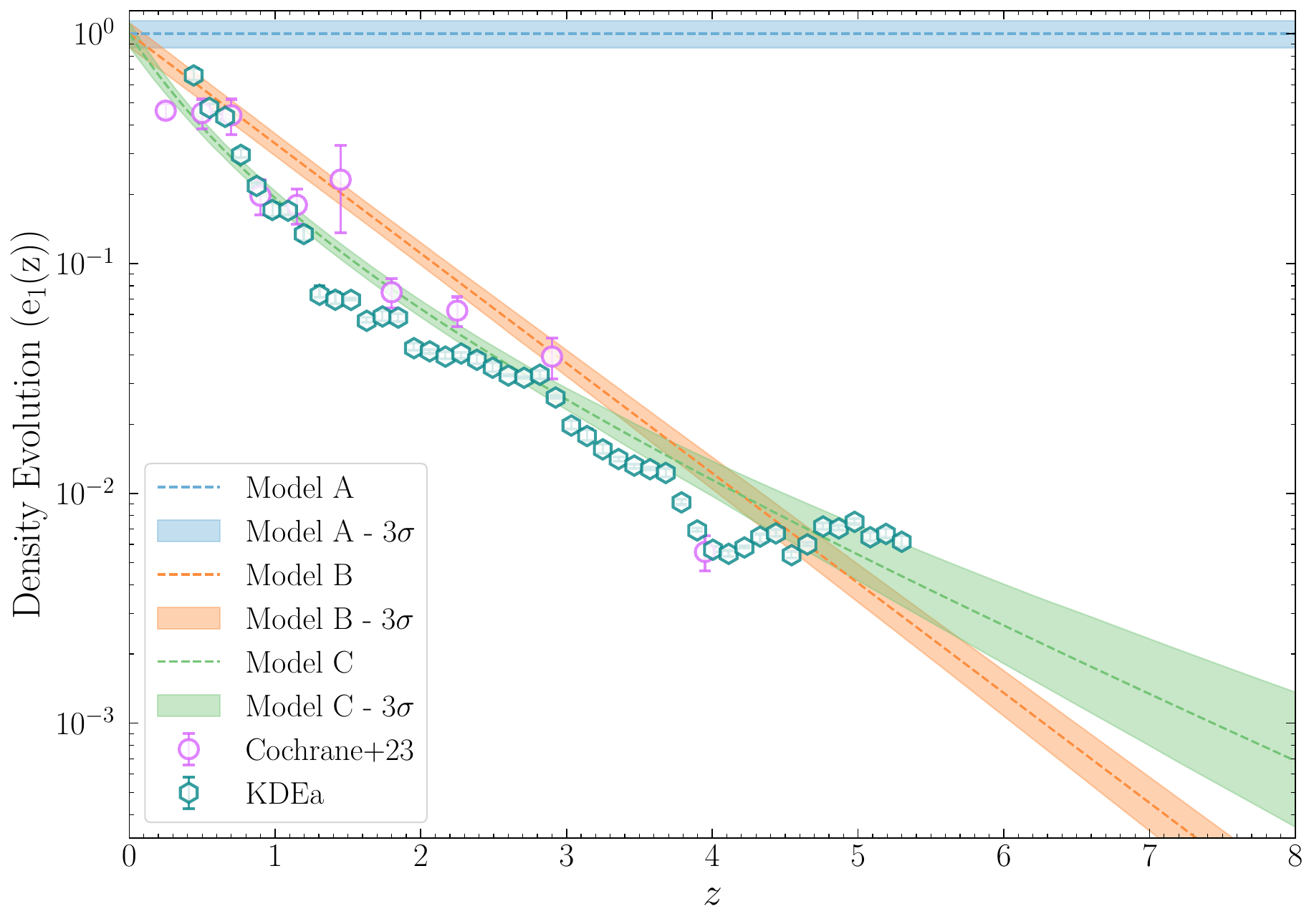}
        \caption {
Redshift evolution of the luminosity (left) and comoving number density (right) of the reference points identified along the KDE-estimated LFs in the ELAIS-N1 field, shown as green hexagons. Assuming a redshift-invariant LF shape, the evolution of these reference points closely traces the underlying LADE trends. Colored curves represent the LE and DE functions derived from our three parametric LF models, with light shaded regions indicating their corresponding $3\sigma$ uncertainty intervals. Purple circles denote the estimates reported by \citet{2023MNRAS.523.6082C}.
        }
        \label{fig:en1evolution}
\end{figure*}

\subsection{Empirical Characterization of the Luminosity Function Evolution with KDE}
\label{sec:empirical_KDE}

As a first step toward constructing a physically motivated parametric model for the SFG LF at low radio frequencies (150\,MHz), we begin by applying the non-parametric KDE method (Section~\ref{sec_methods}) to the ELAIS-N1 field—the deepest of the three fields in our sample. The high completeness and depth of this field allow for a robust, data-driven reconstruction of the LF shape across a wide range of redshifts and luminosities.

This KDE-based estimate serves as an empirical reference for identifying key features of the LF, such as the location of the turnover, the slopes at the faint and bright ends, and possible redshift evolution patterns. Based on these insights, we then propose parametric models that capture the observed LF behavior and proceed to constrain their parameters using the likelihood framework introduced above.

In Figure~\ref{fig:KDE_DELE}, we present the LFs estimated at different redshifts using the adaptive KDE method. Specifically, the redshift range is discretized into a series of grid points, and the LF is computed at each of these redshift slices. The resulting curves are color-coded according to redshift, with the color bar in Figure~\ref{fig:KDE_DELE} indicating the mapping between redshift and color.

As shown in Figure~\ref{fig:KDE_DELE}, the LF exhibits clear evolutionary trends with redshift. As redshift increases, the LF curves systematically shift toward higher luminosities along the horizontal axis. In addition, a systematic decrease in normalization is also observed, with the curves shifting downward along the vertical axis. This pattern suggests that typical SFGs tend to be more luminous at higher redshifts, while their comoving number density decreases over time—consistent with a scenario of LADE.

To quantitatively investigate the trends of density evolution (DE) and luminosity evolution (LE), we identify representative reference points along the KDE-estimated LF curves and track how these points evolve with redshift. Specifically, we compute the first derivative of $\phi_{\rm KDEa}(z, L)$ with respect to $\log_{10} L$, and select the locations where the absolute value of the derivative reaches a minimum. These points, indicated by solid circles in Figure~\ref{fig:KDE_DELE}, correspond to the flattest regions of the $\phi_{\rm KDEa}(z, L)$ curves—i.e., where the LF varies most slowly with luminosity. Their redshift evolution provides additional insight into the combined effects of LADE.

The evolution of these reference points with redshift is visualized in Figure~\ref{fig:en1evolution}, where we track their trajectories in both luminosity (horizontal axis) and comoving number density (vertical axis). This provides a quantitative means of disentangling the respective contributions of LADE to the overall LF evolution. Under the assumption that the shape of the LF remains invariant with redshift, the evolution of these reference points effectively captures the redshift dependence of the LADE functions. Specifically, horizontal shifts reflect luminosity evolution, while vertical displacements correspond to changes in comoving number density. As shown in Figure~\ref{fig:en1evolution}, the trajectories of the reference points reveal clear signatures of both luminosity and density evolution. This empirical evidence motivates the adoption of a mixed-evolution model, in which the LF is parameterized as a combination of redshift-dependent LADE terms.

\subsection{Models for the Luminosity Function of Star-Forming Galaxies}
\label{methods_model}

Guided by the empirical trends revealed in the KDE-based analysis of the ELAIS-N1 field (Section~\ref{sec:empirical_KDE}), we now turn to constructing parametric models for the LF of SFGs. In particular, the observed simultaneous shift in luminosity and normalization with increasing redshift strongly supports a scenario involving both LE and DE.

The SFG LF can be expressed as
\begin{eqnarray}
\label{eq:LF1}
\Phi(z,L) = e_1(z)\, \phi(z=0, L/e_2(z), \eta^j),
\end{eqnarray}
where $e_1(z)$ and $e_2(z)$ denote the redshift-dependent DE and LE functions, respectively, and $\eta^j$ represents the parameters that define the shape of the LF. A constant $\eta^j$ indicates that the LF shape remains invariant with redshift, while a redshift-dependent $\eta^j$ implies luminosity-dependent density evolution \citep[see][for details]{singal2013radio, singal2014gamma}. Consistent with many recent studies \citep[e.g.,][]{2017A&A...602A...5N, 2022ApJ...941...10V, 2024A&A...683A.174W}, we assume that the LF shape does not evolve with redshift, i.e., $\eta^j$ is constant.

Following earlier work \citep[e.g.,][]{smolvcic2009cosmic, Gruppioni_2013, 2022ApJ...941...10V, 2024A&A...683A.174W}, the local LF $\phi(z=0, L/e_2(z=0))$ is modeled using the modified Schechter function proposed by \citet{saunders199060}:
\begin{eqnarray}
\label{eq:LF2}
\begin{aligned}
\phi(z&=0,L/e_2(z=0)) = \frac{dN}{d\log_{10}L} \\
&= \phi_{\star} \left( \frac{L}{L_{\star}} \right)^{1-\beta}
\exp \left[ -\frac{1}{2\gamma^2} \log^2 \left(1 + \frac{L}{L_{\star}} \right) \right],
\end{aligned}
\end{eqnarray}
where $L_{\star}$, $\beta$, $\gamma$, and  $\phi_{\star}$ are free parameters;
$L_{\star}$ marks the characteristic luminosity (or “knee”) of the LF, $\beta$ and $\gamma$ control the slopes at the faint and bright ends, respectively, and $\phi_{\star}$ is the normalization constant.

To capture the observed redshift evolution, we consider three LF models, each adopting the same LE function:
\begin{eqnarray}
\label{e2A}
e_2(z) = (1 + z)^{k_1 + k_2 z},
\end{eqnarray}
where $k_1$ and $k_2$ are free parameters. The DE function $e_1(z)$ differs among models:
\begin{itemize}
\item Model A assumes PLE, commonly used in the literature \citep[e.g.,][]{2017A&A...602A...5N}, with
\begin{eqnarray}
\label{e1A}
e_1(z) = 1.
\end{eqnarray}

\item Model B introduces an exponential density evolution:
\begin{eqnarray}
\label{e1B}
e_1(z) = 10^{p_1 z},
\end{eqnarray}

\item Model C adopts a redshift-dependent power-law form:
\begin{eqnarray}
\label{e1C}
e_1(z) = (1 + z)^{p_1 + p_2 z},
\end{eqnarray}
\end{itemize}
where $p_1$ and $p_2$ are additional free parameters. Models B and C represent the so-called “mixture evolution” or “LADE”  scenarios, which allow for greater flexibility in modeling the joint evolution of number density and luminosity over cosmic time \citep[e.g.,][]{yuan2016mixture, yuan2017mixture, 2010MNRAS.401.2531A}.
In our approach, all parameters in Equation~(\ref{eq:LF2}) are treated as free variables and are fitted simultaneously with the luminosity and density evolution parameters, using a MCMC approach implemented with the {\sc emcee} Python package \citep{foreman2013emcee}. This yields a fully self-consistent determination of both the local LF shape and its cosmic evolution.

\subsection{Model Selection}
\label{methods_select}

To determine which of the proposed LF models (i.e., Models A–C in Section~\ref{methods_model}) provides the best description of the observed data, we apply model selection criteria based on information theory. These statistical tools quantify the trade-off between goodness of fit and model complexity, allowing for an objective comparison of models with different numbers of parameters.

Among them, the Akaike Information Criterion \citep[AIC;][]{1974ITAC...19..716A} is widely used and defined as
\begin{eqnarray}
        \label{aic}
        \text{AIC} = S_\star(\hat{\theta}) + 2q,
\end{eqnarray}
where $S_\star$ is the total negative log-likelihood defined in Equation~(\ref{eq:chi2_2}), $\hat{\theta}$ denotes the maximum-likelihood estimates of the model parameters, and $q$ is the number of free parameters. The model with the lowest AIC value is considered the most favorable in terms of the balance between fit quality and parsimony.

We also consider the Bayesian Information Criterion \citep[BIC;][]{Schwarz1978}, which imposes a stronger penalty for model complexity:
\begin{eqnarray}
        \label{bic}
        \text{BIC} = S_\star(\hat{\theta}) + q \ln n,
\end{eqnarray}
where $n$ is the total number of observed sources. Compared to AIC, BIC tends to prefer simpler models when the sample size is large. In our analysis, both AIC and BIC are used to assess the relative performance of Models A–C.

%The resulting AIC and BIC values for ELAIS-N1 and all fields are presented in Table \ref{en1aicbicpara} and Table \ref{allaicbicpara} respectively. Both criteria consistently favor the LADE model over the PLE model.
%Moreover, for the individual field (e.g. ELAIS-N1), both the AIC and BIC values of Model C are lower than those of Model B, indicating a better fit. In contrast, when considering all fields collectively, Model B yields comparatively lower AIC and BIC values, suggesting it is favored in the joint analysis.

\section{Results}
\label{sec_results}

\begin{figure*}
\centering
\includegraphics[width=\textwidth]{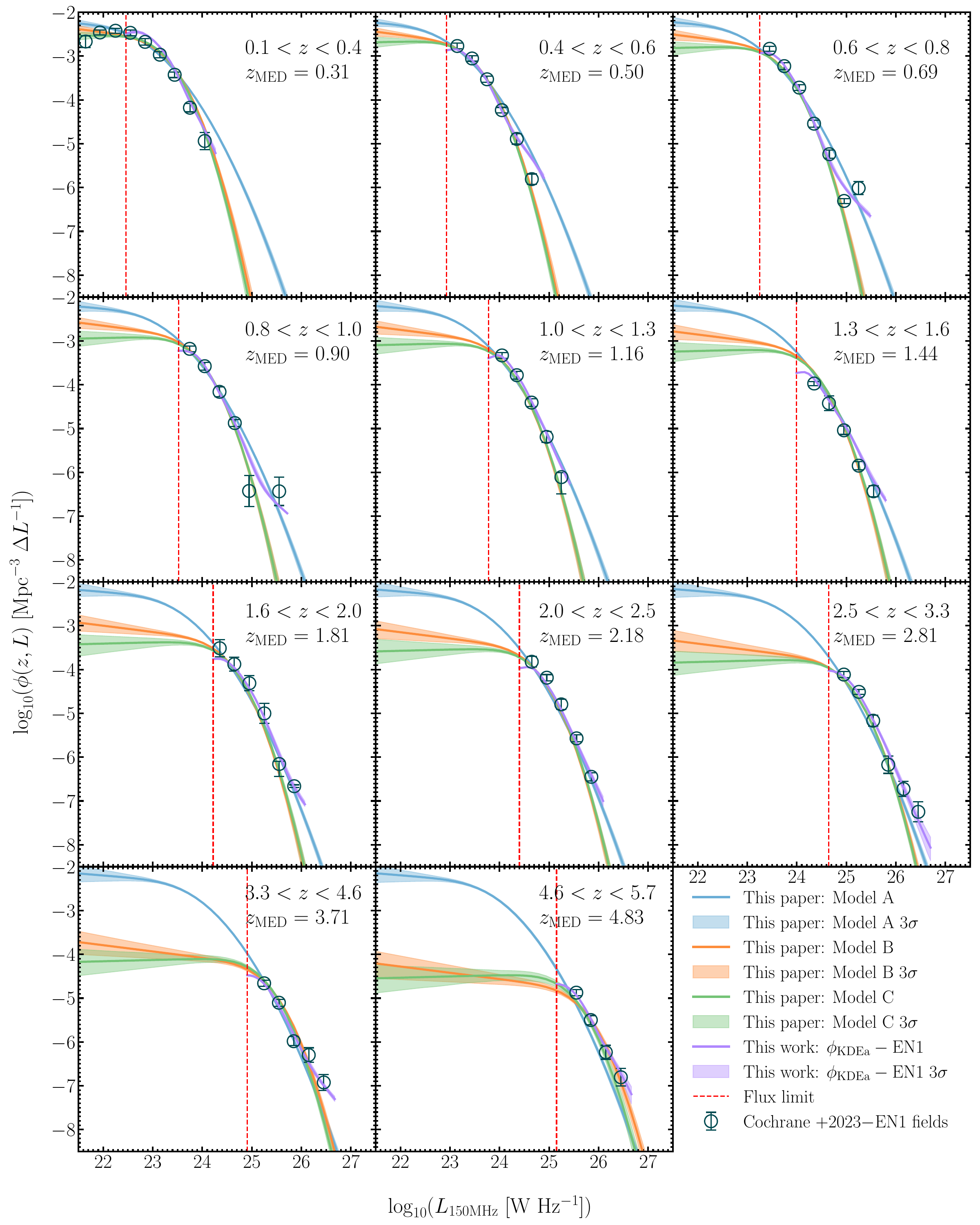}
\caption{
Radio LFs of SFGs in the ELAIS-N1 field at various redshifts. Blue, orange, and green solid lines show the best-fit LFs from Models A, B, and C, respectively. The light shaded area shows the 3$\sigma$ confidence interval. The vertical red dashed line in each panel indicates the luminosity threshold corresponding to the survey flux limit at the given redshift. The solid purple lines indicate adaptive KDE LFs and shaded areas indicate 3$\sigma$ confidence intervals. Circles with error bars denote the binned LF from \cite{2023MNRAS.523.6082C}.
}
\label{fig:en1plf}
\end{figure*}

\begin{figure*}
\centering
\includegraphics[width=\textwidth]{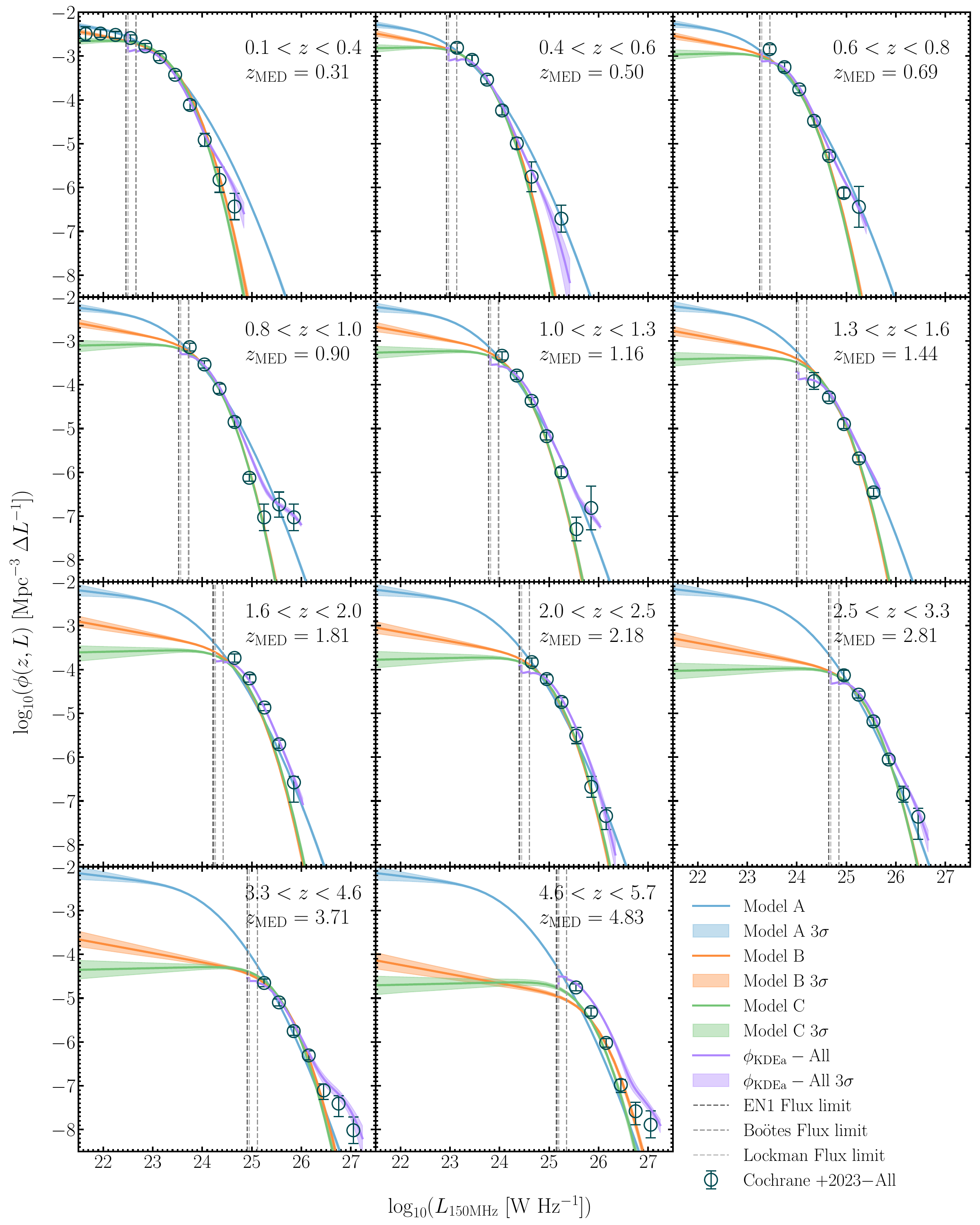}
\caption{
Similar to Figure~\ref{fig:en1plf}, but for the combined analysis of all fields.
The three dashed vertical lines in each panel indicate the luminosity thresholds corresponding to the survey flux limits of the ELAIS-N1, Bo{\"o}tes, and Lockman Hole fields at the given redshift.
}
\label{fig:allplf}
\end{figure*}

\begin{table*}
\centering
\caption{Best-fit parameters for the ELAIS-N1 field: Models A, B, and C}
\centering
\begin{tabular}{lcccccccc}
\hline\hline
Model~~ & $\log_{10}(\phi_{\star})$ & $\log_{10}(L_{\star})$
& $\beta$ & $\gamma$ & $k_1$ & $k_2$ & $p_1$ & $p_2$\\
\hline
A~~ & $-2.295_{-0.018}^{+0.018}$ & $21.718_{-0.061}^{+0.067}$ & $1.089_{-0.041}^{+0.046}$ & $0.707_{-0.007}^{+0.007}$ & $2.571_{-0.026}^{+0.026}$ & $-0.162_{-0.005}^{+0.005}$ & $\ldots$ & $\ldots$\\
B~~ & $-2.423_{-0.018}^{+0.018}$ & $22.283_{-0.036}^{+0.036}$ & $1.142_{-0.026}^{+0.029}$ & $0.427_{-0.008}^{+0.007}$ & $3.640_{-0.029}^{+0.029}$ & $0.002_{-0.007}^{+0.007}$ & $-0.478_{-0.007}^{+0.007}$ & $\ldots$\\
C~~ & $-2.237_{-0.019}^{+0.019}$ & $22.214_{-0.035}^{+0.039}$ & $0.970_{-0.032}^{+0.037}$ & $0.421_{-0.006}^{+0.006}$ & $4.247_{-0.059}^{+0.059}$ & $-0.148_{-0.015}^{+0.015}$ & $-2.239_{-0.083}^{+0.083}$ & $-0.135_{-0.021}^{+0.021}$ \\
\hline
\end{tabular}
{\footnotesize Note. Units --- $\phi_{\star}$: [${\rm Mpc^{-3}\ dex^{-1}}$],\,\, $L_{\star}$: [${\rm W\ Hz^{-1}}$]. The best-fit parameters as well as their 1$\sigma$ errors for models A, B, and C.}
\label{en1modelpara}
\end{table*}

\begin{table*}
\centering
\caption{Best-fit parameters for All fields: Models A, B, and C}
\centering
\begin{tabular}{lcccccccc}
\hline\hline
Model~~ & $\log_{10}(\phi_{\star})$ & $\log_{10}(L_{\star})$
& $\beta$ & $\gamma$ & $k_1$ & $k_2$ & $p_1$ & $p_2$\\
\hline
A~~ & $-2.381_{-0.013}^{+0.013}$ & $21.784_{-0.040}^{+0.045}$ & $1.135_{-0.028}^{+0.031}$ & $0.703_{-0.004}^{+0.004}$ & $2.791_{-0.017}^{+0.017}$ & $-0.186_{-0.003}^{+0.003}$ & $\ldots$ & $\ldots$\\
B~~ & $-2.604_{-0.013}^{+0.013}$ & $22.555_{-0.022}^{+0.022}$ & $1.209_{-0.016}^{+0.016}$ & $0.384_{-0.005}^{+0.005}$ & $3.914_{-0.020}^{+0.020}$ & $-0.034_{-0.005}^{+0.005}$ & $-0.487_{-0.004}^{+0.004}$ & $\ldots$\\
C~~ & $-2.324_{-0.013}^{+0.013}$ & $22.312_{-0.024}^{+0.024}$ & $0.974_{-0.023}^{+0.023}$ & $0.382_{-0.004}^{+0.004}$ & $4.670_{-0.033}^{-0.033}$ & $-0.204_{-0.007}^{+0.007}$ & $-2.536_{-0.047}^{+0.047}$ & $-0.094_{-0.011}^{+0.011}$\\
\hline
\end{tabular}
{\footnotesize Note. Units --- $\phi_{\star}$: [${\rm Mpc^{-3}\ dex^{-1}}$],\,\, $L_{\star}$: [${\rm W\ Hz^{-1}}$]. The best-fit parameters as well as their 1$\sigma$ errors for models A, B, and C.}
\label{allmodelpara}
\end{table*}

\subsection{Luminosity Function Fitting for the the ELAIS-N1 field}
\label{result_LFs}
Figure~\ref{fig:en1plf} presents the best-fitting LFs for the ELAIS-N1 field derived from Model A (blue solid line), Model B (orange solid line), and Model C (green solid line). All LFs are evaluated at the rest-frame frequency of 150\,MHz. For comparison, we also include the binned LFs from \citet{2023MNRAS.523.6082C}, shown as blue circles with error bars. The KDE-based non-parametric estimates are shown as purple solid lines, with the shaded regions indicating the $3\sigma$ confidence intervals, and are in good agreement with the binned LFs from \citet{2023MNRAS.523.6082C}.

For the ELAIS-N1 field, the marginalized one- and two-dimensional posterior distributions of the model parameters are shown in Figures~\ref{fig:en1cornerplotA}, \ref{fig:en1cornerplotB}, and \ref{fig:en1cornerplotC} for Models A, B, and C, respectively. These corner plots demonstrate that all parameters are well constrained across the three models. Table~\ref{en1modelpara} summarizes the best-fitting parameter values and their $1\sigma$ uncertainties for the ELAIS-N1 field.

Notably, both the binned LFs from \citet{2023MNRAS.523.6082C} and our KDE-based non-parametric estimates exhibit a distinct bump at the high-luminosity end. Because the non-parametric estimates are entirely data-driven, this excess directly reflects the structure of the observed sample rather than assumptions in the modeling. When compared with the best-fitting parametric LFs, the non-parametric estimates show systematically higher number densities at the bright end, indicating that the excess arises relative to the model predictions. While low-number statistics could in principle contribute to this feature, it is unlikely to be the dominant factor, as a similar bright-end excess is seen across nearly all redshift bins in Figure~\ref{fig:en1plf}. To assess the possible influence of these sparse data points, we repeated the parametric fitting after excluding the highest-luminosity bin in each redshift interval—typically containing only one or two sources. The resulting best-fit parameters and evolutionary trends changed negligibly, confirming that our conclusions are robust against the inclusion of these bins. A plausible explanation for this feature is the contamination from misclassified sources: although our sample is selected to represent SFGs, it is likely that some AGNs remain unidentified and contribute disproportionately at high luminosities. The presence of such AGNs would elevate the number density in the bright end, thereby producing the observed bump.

In each panel of Figure~\ref{fig:en1plf}, we indicate the luminosity threshold corresponding to the survey flux limit of the ELAIS-N1 field at the given redshift by a vertical red dashed line.
Above this threshold, Models~B and C yield nearly identical results and show good agreement with both the binned LFs and the KDE-based estimates. Model~A, in contrast, exhibits a slightly shallower decline at the bright end, deviating mildly from the other two models. This behavior appears to reflect a stronger response to the mild upturn observed at the high-luminosity end of the binned and KDE-based LFs—possibly caused by residual AGN contamination—suggesting that Model~A may be overfitting to this bright-end feature.

Below the luminosity threshold, the three models begin to diverge more significantly. Since this regime lies beyond the reach of the observed data, comparisons with the binned LFs or KDE estimates are not feasible. In such extrapolated domains, statistical model selection criteria become essential for assessing model performance. To this end, we employ the AIC and the BIC both of which balance goodness-of-fit against model complexity. Table~\ref{en1aicbicpara} presents the AIC and BIC values for Models~A, B, and C in the ELAIS-N1 field. Model~C yields the lowest values in both criteria, followed closely by Model~B, while Model~A is strongly disfavored.

To highlight the relative performance, we also compute $\Delta \mathrm{AIC}$ and $\Delta \mathrm{BIC}$ with respect to the best-performing model. A difference larger than 10 is typically interpreted as strong evidence against a model. In our case, not only is Model~A clearly ruled out, but Model~C is also decisively preferred over Model~B. Model~B also performs significantly better than Model~A, suggesting that incorporating some form of density evolution improves the model fit. These results support the adoption of a mixed-evolution scenario—combining both luminosity and density evolution—as the most appropriate framework for modeling the SFG LF in the ELAIS-N1 field.

In addition to the statistical model comparisons, we also examine how well the fitted evolution trends from Models~A, B, and C reproduce the non-parametric estimates of LADE obtained from the KDE analysis. As shown in Figure~\ref{fig:en1evolution}, Model~C provides the closest match to the KDE-derived LE and DE trends, while Model~B also shows broadly consistent behavior within uncertainties. In contrast, Model~A deviates more noticeably, particularly in the density evolution component. This consistency between the non-parametric and parametric results provides further support for adopting a mixed-evolution framework when modeling the SFG LF.

\begin{table}
\centering
\caption{Model comparison based on AIC and BIC for the ELAIS-N1 field. $\Delta$AIC and $\Delta$BIC are computed relative to the model with the minimum value.}
\begin{tabular}{lcccc}
\hline\hline
Model~~ & AIC & $\Delta$AIC & BIC & $\Delta$BIC \\
\hline
A~~ & 387621.2 & 1709.8 & 387669.1 & 1693.8 \\
B~~ & 386020.7 & 109.3  & 386076.6 & 101.3  \\
C~~ & 385911.4 & 0.0    & 385975.3 & 0.0    \\
\hline
\end{tabular}
\label{en1aicbicpara}
\end{table}

\begin{table}
\centering
\caption{AIC and BIC values for all fields: Models A, B, and C. Differences $\Delta \mathrm{AIC}$ and $\Delta \mathrm{BIC}$ are calculated relative to the best-performing model (Model B).}
\begin{tabular}{lcccc}
\hline\hline
Model~~ & AIC & $\Delta \mathrm{AIC}$ & BIC & $\Delta \mathrm{BIC}$ \\
\hline
A~~ & 1038437.2 & 15020.3 & 1038490.9 & 15011.4 \\
B~~ & 1023417.0 & 0.0     & 1023479.5 & 0.0     \\
C~~ & 1028178.9 & 4761.9  & 1028250.4 & 4770.9  \\
\hline
\end{tabular}
\label{allaicbicpara}
\end{table}

\begin{figure*}[ht]
        \centering
        \includegraphics[width=0.5\textwidth]{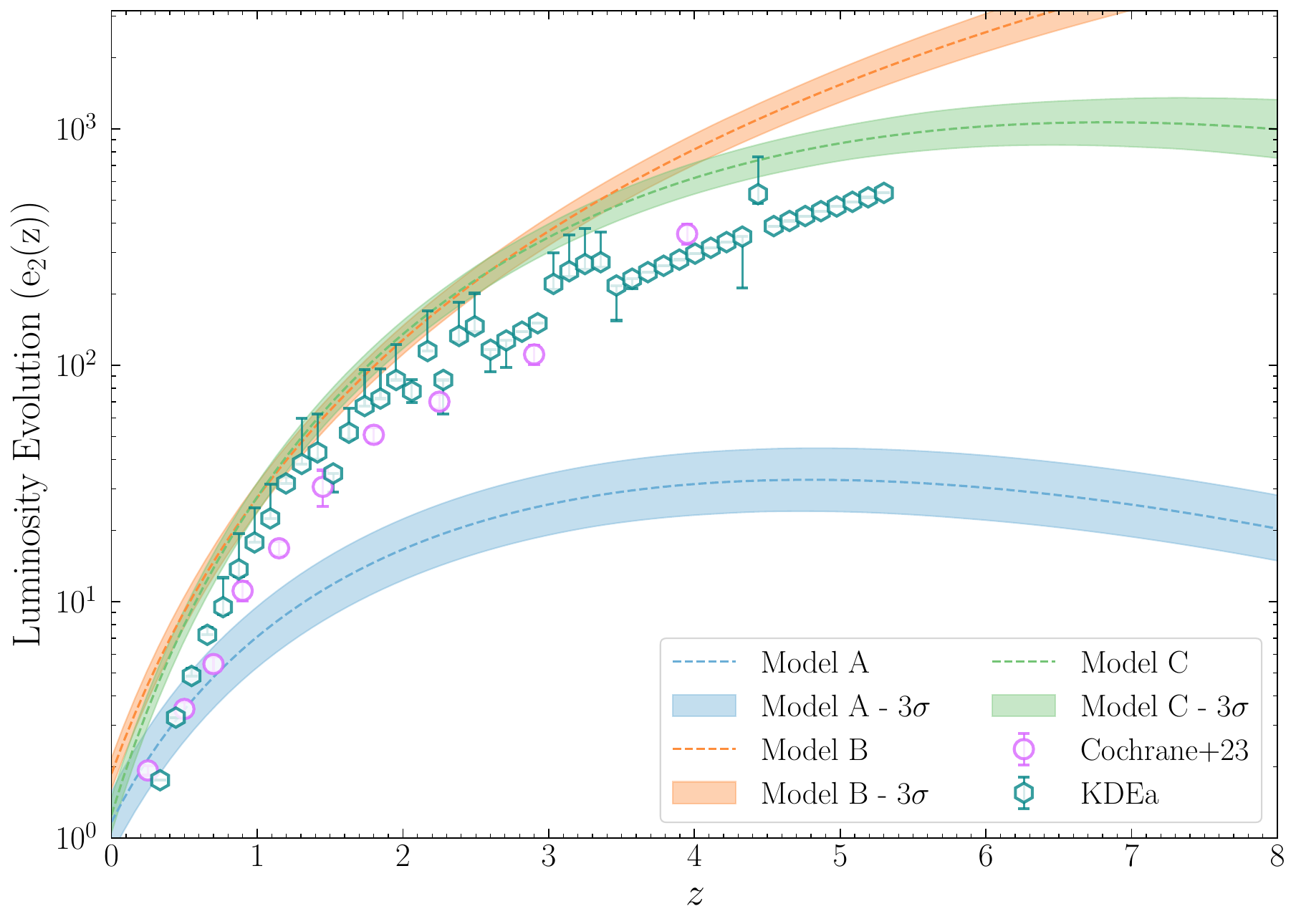}
        \hspace{-0.1in}
        \includegraphics[width=0.5\textwidth]{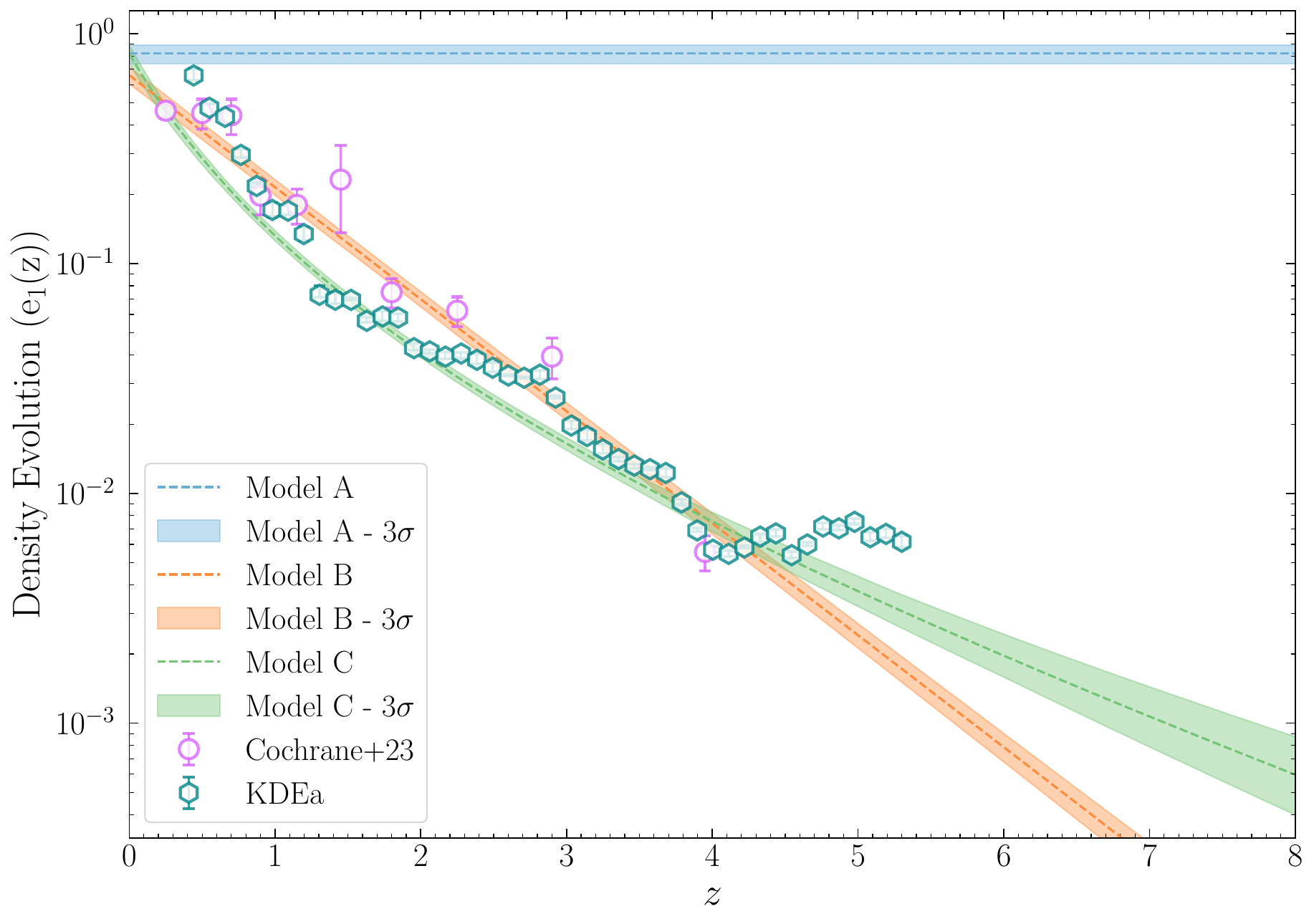}
\caption{
LE and DE fitted from our three parametric LF models using the combined sample of all fields, with light shaded regions indicating the corresponding $3\sigma$ uncertainty intervals. 
The green hexagons and purple circles represent the LE and DE trends based solely on the ELAIS-N1 field (same as in Figure~\ref{fig:en1evolution}).
}

        \label{fig:allevolution}
\end{figure*}

\subsection{Luminosity Function Fitting for the Combined Sample}

To further constrain the SFG LF, we combine the data from the three fields—ELAIS-N1, Bo{\"o}tes, and Lockman Hole—into a unified sample and apply the same modeling framework as in the individual field analysis. The best-fitting LFs obtained using Models~A, B, and C are compared against the averaged KDE-based non-parametric estimates derived from the individual fields. The marginalized posterior distributions for Models~A, B, and C in the combined analysis are presented in Figures~\ref{fig:allcornerplotA}, \ref{fig:allcornerplotB}, and \ref{fig:allcornerplotC}, respectively, showing that all model parameters are well constrained. The corresponding best-fitting values and their $1\sigma$ uncertainties are summarized in Table~\ref{allmodelpara}.

Figure~\ref{fig:allplf} presents the resulting LFs from the combined analysis. The best-fit LFs for Models~A (blue), B (orange), and C (green) are shown in each redshift bin, alongside the binned estimates from \citet{2023MNRAS.523.6082C} and the averaged KDE results (purple solid lines with $3\sigma$ confidence intervals). The three vertical dashed lines in each panel indicate the luminosity thresholds corresponding to the survey flux limits of the ELAIS-N1, Bo{\"o}tes, and Lockman Hole fields at the given redshift. As in the ELAIS-N1 field, both the binned and KDE LFs exhibit a mild bright-end excess across redshift bins, potentially due to residual AGN contamination. The overall behavior of the three parametric models resembles that observed in the ELAIS-N1 field: they agree above the luminosity thresholds, while more substantial differences emerge at the faint end, where observational constraints are absent.

Figure~\ref{fig:allplf} presents the resulting LFs from the combined analysis. The best-fit LFs for Models~A (blue), B (orange), and C (green) are shown in each redshift bin, alongside the binned estimates from \citet{2023MNRAS.523.6082C} and the averaged KDE results (purple solid lines with $3\sigma$ confidence intervals). Overall, the KDE LFs are in good agreement with the binned estimates of \citet{2023MNRAS.523.6082C} across most redshift bins. However, at the highest redshift interval ($4.6 < z < 5.7$), a noticeable deviation appears. This arises partly because the original KDE formalism \citep{yuan2022flexible} is designed for a single flux-limited sample and cannot directly account for multiple surveys with different flux limits. In our implementation, we therefore compute the KDE LFs separately for the three fields and take their average as an approximate combined estimate. While this provides a practical compromise, it inevitably introduces additional uncertainty at the high-redshift, high-luminosity end, where the data are sparse. The three vertical dashed lines in each panel indicate the luminosity thresholds corresponding to the survey flux limits of the ELAIS-N1, Bo{\"o}tes, and Lockman Hole fields at the given redshift. As in the ELAIS-N1 field, both the binned and KDE LFs exhibit a mild bright-end excess across redshift bins, potentially due to residual AGN contamination. The overall behavior of the three parametric models resembles that observed in the ELAIS-N1 field: they agree above the luminosity thresholds, while more substantial differences emerge at the faint end, where observational constraints are absent.

Figure~\ref{fig:allevolution} shows the fitted LE and DE functions from Models~A, B, and C using the combined dataset, overlaid with the KDE-derived LE and DE trends (green hexagons, same as Figure~\ref{fig:en1evolution}) based solely on the ELAIS-N1 field. For Models~B and~C, the fitted LE and DE functions broadly reproduce the KDE-derived evolutionary trends, although the LE curves rise more steeply and the DE curves decline more rapidly than those inferred from the ELAIS-N1 field alone. Interestingly, this behavior reveals a possible “see-saw” degeneracy between LE and DE: when LE evolves more strongly with redshift, DE tends to decrease more steeply, and vice versa. This degeneracy likely reflects the intrinsic trade-off between LE and DE in modeling the redshift–luminosity distribution of SFGs.

Table~\ref{allaicbicpara} summarizes the AIC and BIC values for the three models. The $\Delta \mathrm{AIC}$ and $\Delta \mathrm{BIC}$ values are computed relative to the best-performing model, Model~B. The results show that Model~A, which assumes PLE, is strongly disfavored. Although Model~C provides greater flexibility through an additional density evolution parameter, it does not outperform the simpler Model~B in the joint analysis. These findings reinforce the importance of including density evolution in modeling the SFG LF, while also highlighting the balance between model complexity and data support: a simpler mixed-evolution model (Model~B) offers the best overall fit when all fields are considered.

Interestingly, while Model~C is statistically preferred in the ELAIS-N1 field, the combined analysis across all three fields identifies Model~B as the optimal model based on both AIC and BIC. This shift can be understood in terms of the trade-off between model complexity and generalizability. The additional degree of freedom in Model~C may provide improved flexibility in fitting detailed LF features in deep fields such as ELAIS-N1, which probe wider luminosity and redshift ranges. However, when applied to the combined dataset—including the shallower Bo{\"o}tes and Lockman Hole fields—this flexibility offers diminishing returns, as these two fields do not significantly extend the luminosity coverage nor enhance constraints near the LF knee where model differences are most evident.

Although the inclusion of multiple fields increases the total number of sources, it does not necessarily improve constraints on the critical regions of the LF. Instead, the shallow fields primarily contribute data in luminosity ranges where all models already converge, while still increasing the penalty term in AIC and BIC due to added complexity. Consequently, the statistical advantage of Model~C becomes diluted, and the simpler Model~B emerges as the more robust and generalizable choice for the full sample.

These results emphasize an important methodological consideration: combining datasets with varying depths and flux limits is only beneficial when the added fields offer complementary constraints. In this study, the Bo{\"o}tes and Lockman Hole fields do not augment the high-luminosity end or better constrain the turnover of the LF; hence, their inclusion does not enhance model discrimination and may even obscure subtle evolutionary features detectable in the deeper ELAIS-N1 field.

In summary, Model~C provides the best fit to the deepest field, capturing more detailed redshift evolution features, but Model~B delivers a more parsimonious and statistically favored description when all fields are considered. We thus conclude that while Model~C may be preferable in deep-field studies with extensive redshift coverage, Model~B offers greater stability and generalizability across heterogeneous survey conditions.

\subsection{Comparing ELAIS-N1 and Combined Fields: Role of External Constraints}
\label{result_sc}

To evaluate the impact of the additional constraints introduced in Equation~(\ref{eq:chi2_2}), we now examine how the best-fit models reproduce the observed LRLF and Euclidean-normalized SCs. We note that the observed LRLF and source counts were also included as constraints in the  fitting procedure (see Section~\ref{sec_methods}). Therefore, the comparisons presented here should not be regarded  as independent tests, but rather as consistency checks to illustrate how well the fitted models  reproduce the basic observational quantities on which they are based.

\begin{figure*}
	\centering
    \includegraphics[width=0.5\textwidth]{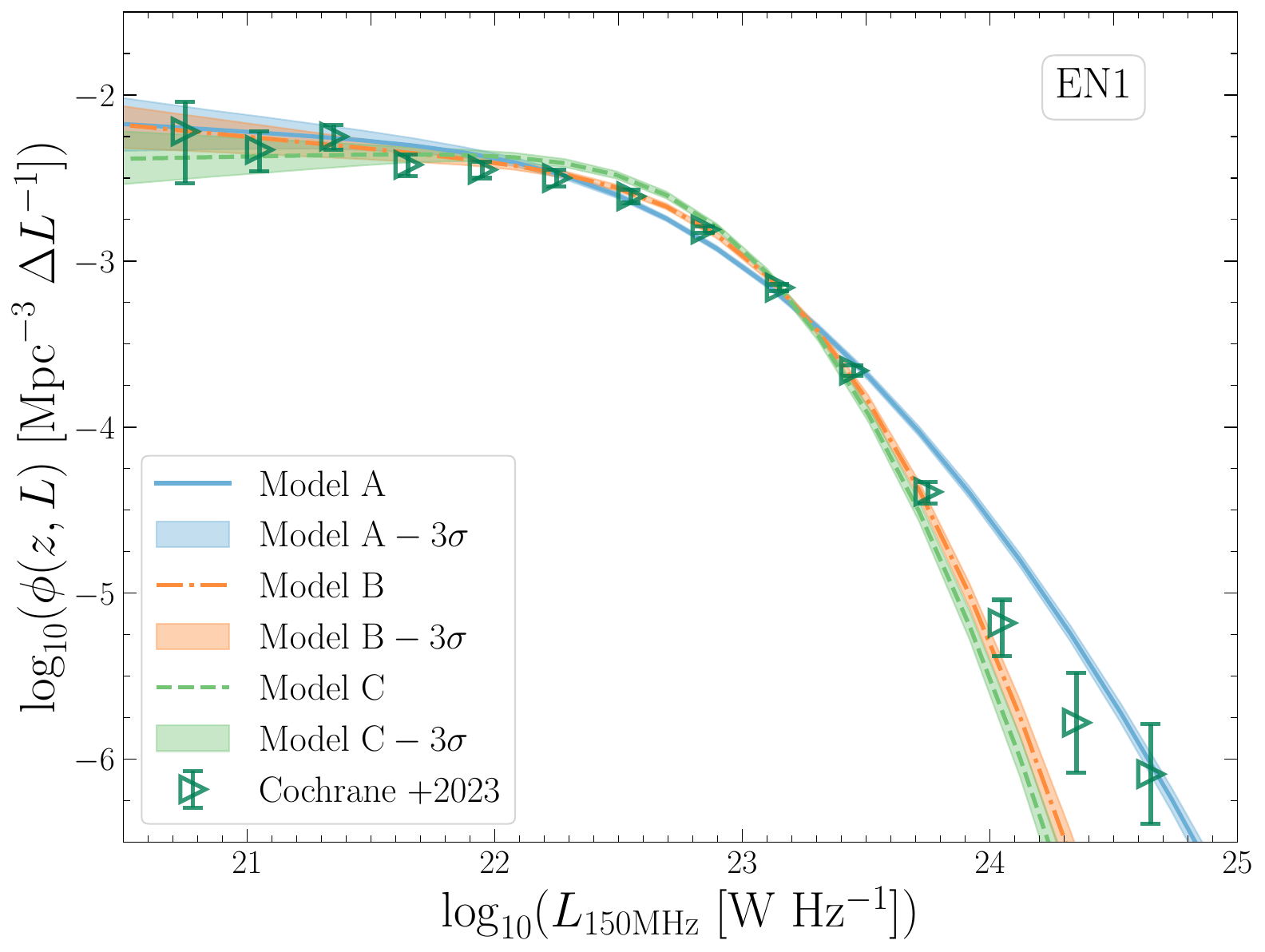}
	\hspace{-0.1in}
    \includegraphics[width=0.5\textwidth]{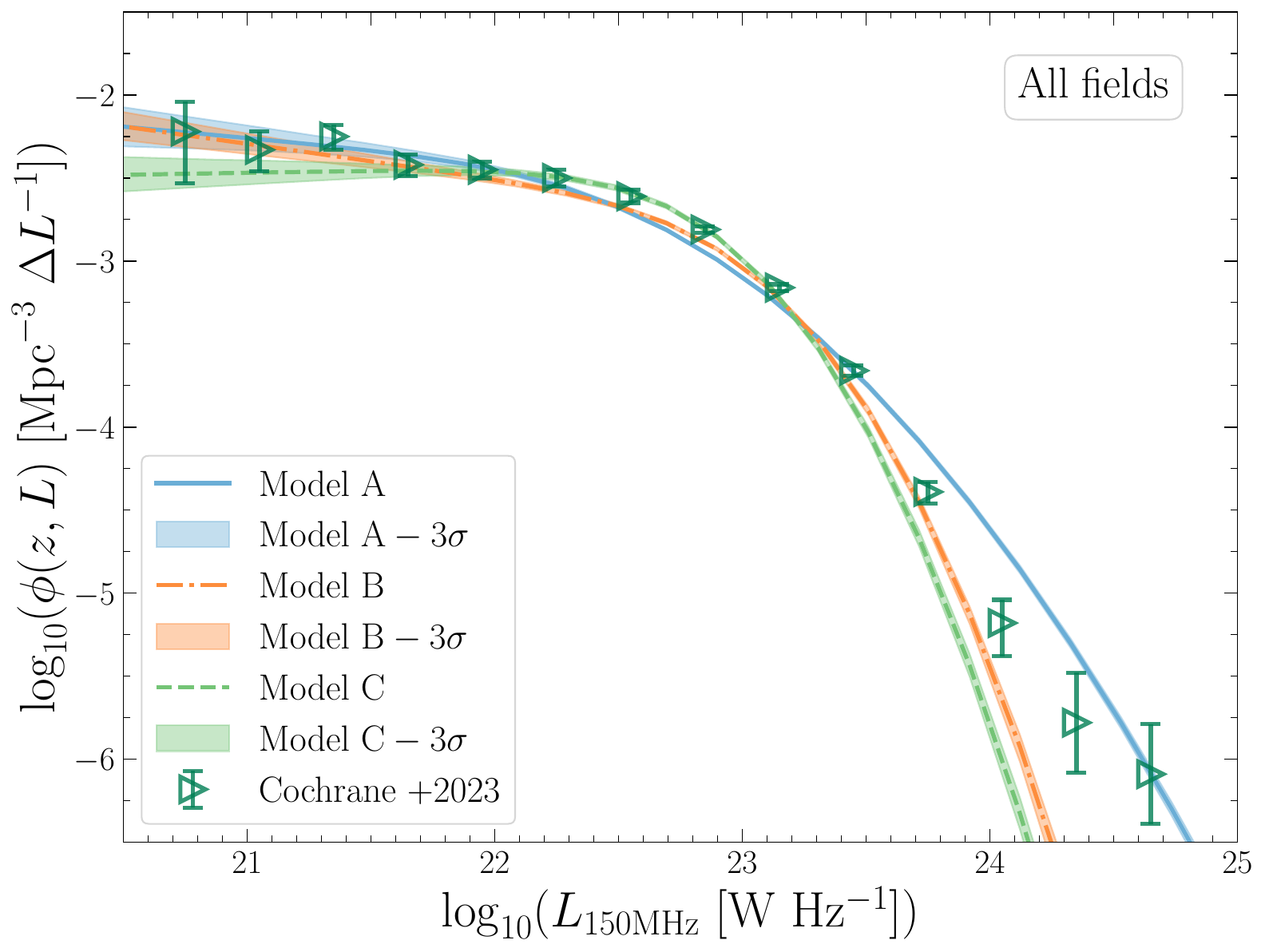}
	\caption{
    Local radio LF at 150 MHz of SFGs for the ELAIS-N1 field (\textit{left}) and the combined sample of all fields (\textit{right}). In both panels, the right-pointing triangles with error bars represent the binned LRLF measurements from \citet{2023MNRAS.523.6082C}, while the colored curves show the corresponding best-fit model predictions from this work.}
	\label{fig:LRLF}
\end{figure*}

Figure~\ref{fig:LRLF} provides a direct comparison between the modeled and observed LRLFs at 150\,MHz for both the ELAIS-N1 field (left panel) and the combined sample of all fields (right panel). In both cases, the right-pointing triangles represent the binned LRLF measurements from \citet{2023MNRAS.523.6082C}, while the colored curves show the best-fit predictions from Models~A (blue), B (orange), and C (green). Models~B and C show excellent agreement with the observed LRLF in the range $L \lesssim 10^{24}\,\mathrm{W\,Hz^{-1}}$.
However, in the two highest-luminosity bins, the binned LRLF shows a noticeable excess that is not captured by Models~B or C. This feature likely reflects residual AGN contamination, or the presence of galaxies exhibiting a coexistence of AGN activity and star formation. Such hybrid systems—where both star formation and AGN contribute significantly to the radio emission—are known to be common in the intermediate-to-high luminosity regime, particularly in composite galaxies or systems hosting low-excitation radio AGN \citep[e.g.,][]{2014MNRAS.440..269M,2012ApJ...745..172D,Delvecchio_2017,2018MNRAS.475.3010G}. Model~A, which predicts a shallower decline at the bright end, appears to overfit the observed excess in an attempt to accommodate it. However, this results in a poorer overall fit across the full luminosity range.

\begin{figure*}
	\centering \includegraphics[width=0.5\textwidth]{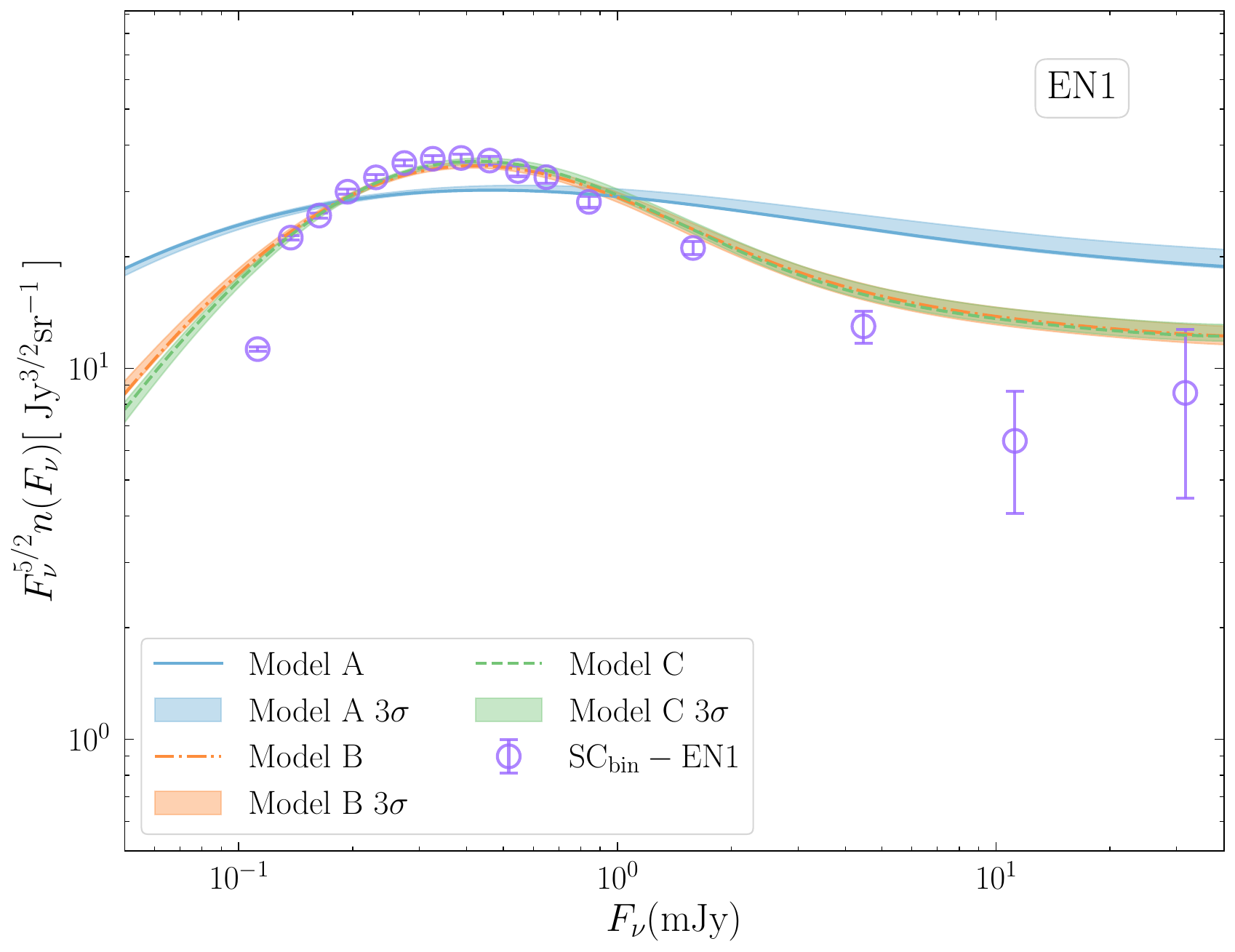}
	\hspace{-0.1in} \includegraphics[width=0.5\textwidth]{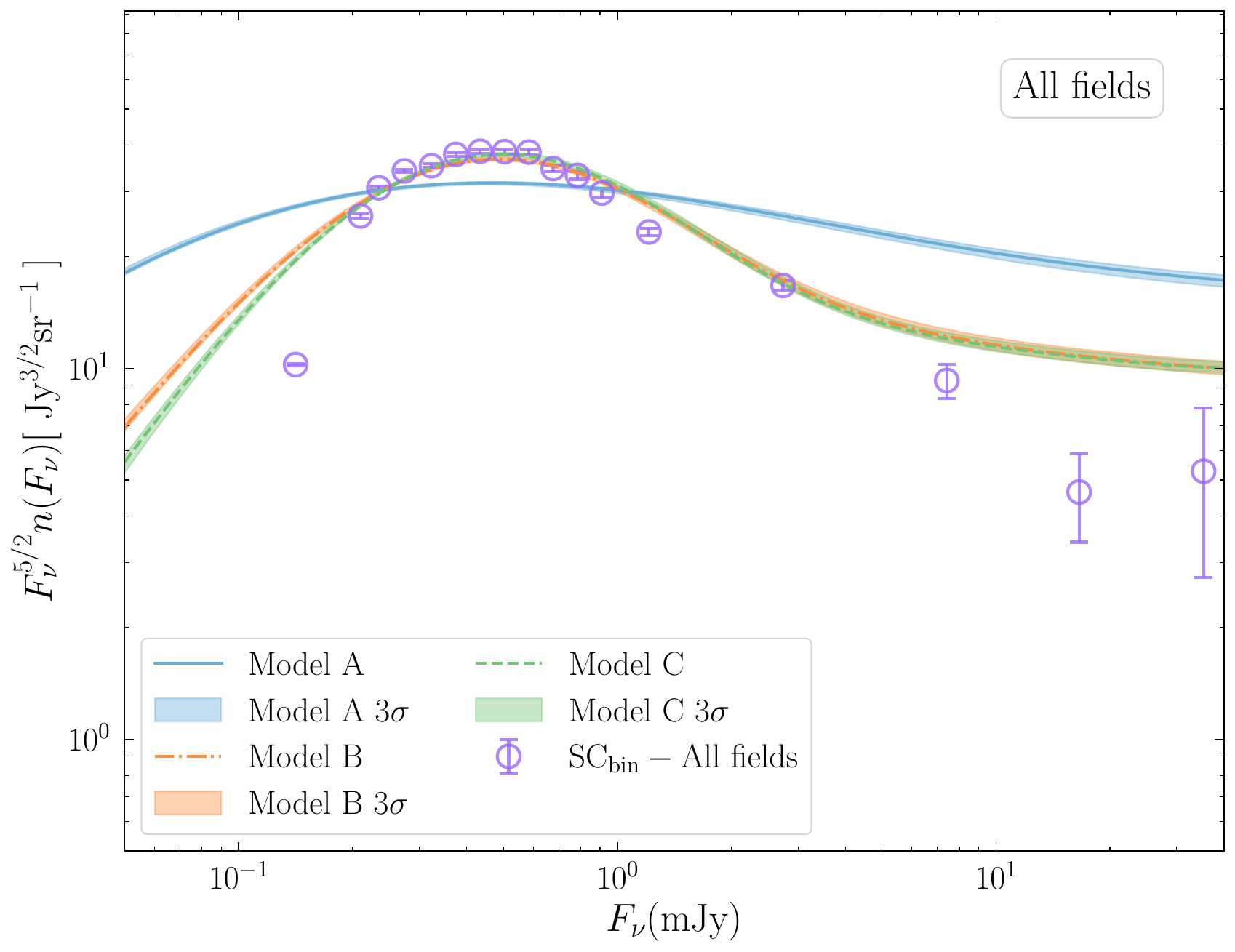}
	\caption{
     Comparison of our best-fit models with the SCs obtained using the binned method in the ELAIS-N1 field (left panel) and combined all three fields (right panel). In both panels, the blue solid line, orange dashed-dotted line, and green dashed line represent the best-fit SCs for Models A, B, and C, respectively. The purple circles denote the SCs obtained using the bin method.
    }
	\label{fig:sc}
\end{figure*}

Figure~\ref{fig:sc} presents a comparison between the observed and modeled Euclidean-normalized radio SCs. The left panel corresponds to the ELAIS-N1 field, while the right panel shows results from the combined dataset of all three fields. In both panels, the purple circles represent the binned counts derived from the LOFAR flux density distributions using the binning method, and the colored lines indicate the predictions from our three parametric models. The blue solid, orange dash-dotted, and green dashed lines correspond to Models~A, B, and C, respectively. Shaded regions indicate the $3\sigma$ confidence intervals.

Models~B and C reproduce the observed SCs well over a wide range of flux densities, for both the ELAIS-N1 field and the full combined sample. In contrast, Model~A shows substantial deviations across the entire flux range, highlighting its overall poorer agreement with the observed SCs. It is also worth noting that at the bright end ($F_\nu \gtrsim 10\,\mathrm{mJy}$), the increased scatter in the binned SCs likely arises from a combination of Poisson noise and classification uncertainties. In this regime, the separation between SFGs and AGNs becomes increasingly ambiguous: some AGNs may be misclassified as SFGs, while certain high-luminosity SFGs exhibiting AGN-like features might be incorrectly excluded. This dual source of contamination, combined with low-number statistics, makes the bright-end flux range particularly challenging to model reliably. In addition, at the faint end, the binned SCs drop significantly below the model predictions. This underestimation likely arises from completeness issues near the flux limit in the three fields, where faint sources are more easily missed, leading to a downward bias in the SCs.

In summary, both Model~B and Model~C consistently reproduce the observed LRLF and Euclidean-normalized SCs. Their success in matching the observational data demonstrates the robustness and flexibility of the mixed-evolution framework in capturing the redshift and luminosity trends of SFGs. In contrast, Model~A, which assumes pure luminosity evolution, fails to provide an adequate fit to either constraint. These results further support the conclusion that incorporating density evolution is essential for accurately modeling the radio LF of SFGs.

\section{Discussion and Conclusions}
\label{sec_discussion}
We have developed a non-parametric framework for characterizing the evolution of the 150\,MHz luminosity function (LF) of star-forming galaxies (SFGs). By applying kernel density estimation (KDE) in the luminosity--redshift plane, we reconstruct a smooth estimate of the joint source distribution that enables evaluation of the LF at arbitrary redshifts, without the need for binning or a predefined analytic form.

Building on this non-parametric foundation, we examine the LF evolution by tracking the displacement of key reference points along the KDE-derived curves, providing a direct, data-driven view of luminosity and density evolution (LADE). These empirical trends then serve as guidance for constructing parametric models, which are fitted using a global maximum-likelihood approach. This combination links the empirical flexibility of KDE with the statistical rigor of parametric inference, providing a unified view of the evolving radio SFG population.

The evolution of the LF has also been studied in \citet{2023MNRAS.523.6082C} using deep 150\,MHz data from the Low Frequency Array (LOFAR) Two-metre Sky Survey (LoTSS). In their analysis, the redshift dependence of the LF parameters was derived by dividing the sample into redshift bins and constructing binned LFs with the $1/V_{\mathrm{max}}$ method. These LF points were then fitted with a parametric form matching the locally derived LF, with the faint-end slope $\alpha$ and high-luminosity cut-off $\sigma$ fixed, while $L_*$ and $\phi_*$ were allowed to vary. This yielded a pair of $(L_*, \phi_*)$ values per redshift slice, forming the basis of their evolutionary trends.

Our parametric modeling follows the same general assumption as \citet{2023MNRAS.523.6082C}, namely that the LF shape is fixed with redshift and the evolution is captured through changes in $L_*$ and $\phi_*$. The main difference lies in how these trends are inferred: rather than fitting $(L_*, \phi_*)$ independently in redshift bins, we perform a global maximum-likelihood fit across all fields and redshifts, guided by the non-parametric KDE results.

In this context, our analysis extends the methodology of \citet{2023MNRAS.523.6082C} by integrating non-parametric reconstruction and parametric modeling into a single, data-driven framework. This reduces sensitivity to binning choices and allows continuous evolution to emerge directly from the data. The main findings of this work are summarized below.

\begin{enumerate}
\item The KDE analysis reveals clear empirical evidence for simultaneous luminosity and density evolution (LADE): the LF systematically shifts toward higher luminosities and lower normalizations with increasing redshift.

\item Guided by the KDE trends, we developed three parametric LF models: Model~A (pure luminosity evolution, PLE), and Models~B and~C (both LADE forms differing in their density evolution). These models were fitted using a global maximum-likelihood framework that incorporates completeness corrections and observational constraints from the local LF and Euclidean-normalized source counts.

\item Models~B and~C successfully reproduce the observed LFs and source counts across a broad luminosity and flux range, confirming the need for models including density evolution. Model~A performs significantly worse, highlighting the limitations of pure luminosity evolution. Model~C provides the best fit for the deepest field (ELAIS-N1), while the simpler Model~B yields the most statistically favored and stable results in the combined-field analysis.

\item A mild excess persists at the bright end of the LF across different estimators (binned and KDE). This feature likely reflects residual active galactic nuclei (AGN) contamination rather than a genuine SFG population, emphasizing the need for improved AGN/SFG separation in future radio surveys.

\item The combined use of non-parametric KDE and parametric maximum-likelihood modeling offers a flexible and statistically robust framework for tracing the cosmic evolution of radio-selected SFGs. With future surveys such as the Square Kilometre Array (SKA) vastly expanding the accessible ranges in luminosity and redshift, this approach will be essential for fully exploiting the scientific potential of next-generation radio observations.

\end{enumerate}

\begin{acknowledgments}
We thank the anonymous reviewer for the many constructive comments and suggestions, leading to a clearer description of these results. We acknowledge financial support from the Science Fund for Distinguished Young Scholars of Hunan Province (Grant No. 2024JJ2040), the National Natural Science Foundation of China (Grant Nos. 12073069, 12075084, 12275080, and 12393813), the Major Basic Research Project of Hunan Province (Grant No. 2024JC0001), and the Innovative Research Group of Hunan Province (Grant No. 2024JJ1006). Z.Y. is supported by the Xiaoxiang Scholars Programme of Hunan Normal University. We thank R.~K.~Cochrane for insightful discussions and valuable guidance on the data classification process. 
LOFAR data products were provided by the LOFAR Surveys Key Science project (LSKSP; https://lofar-surveys.org/) and were derived from observations with the International LOFAR Telescope (ILT). LOFAR (van Haarlem et al. 2013) is the Low Frequency Array designed and constructed by ASTRON. It has observing, data processing, and data storage facilities in several countries, which are owned by various parties (each with their own funding sources), and which are collectively operated by the ILT foundation under a joint scientific policy. The efforts of the LSKSP have benefited from funding from the European Research Council, NOVA, NWO, CNRS-INSU, the SURF Co-operative, the UK Science and Technology Funding Council and the Jülich Supercomputing Centre.
\end{acknowledgments}

\FloatBarrier  % 推荐
\appendix

\begin{figure*}
        \centering
        \includegraphics[width=\textwidth]{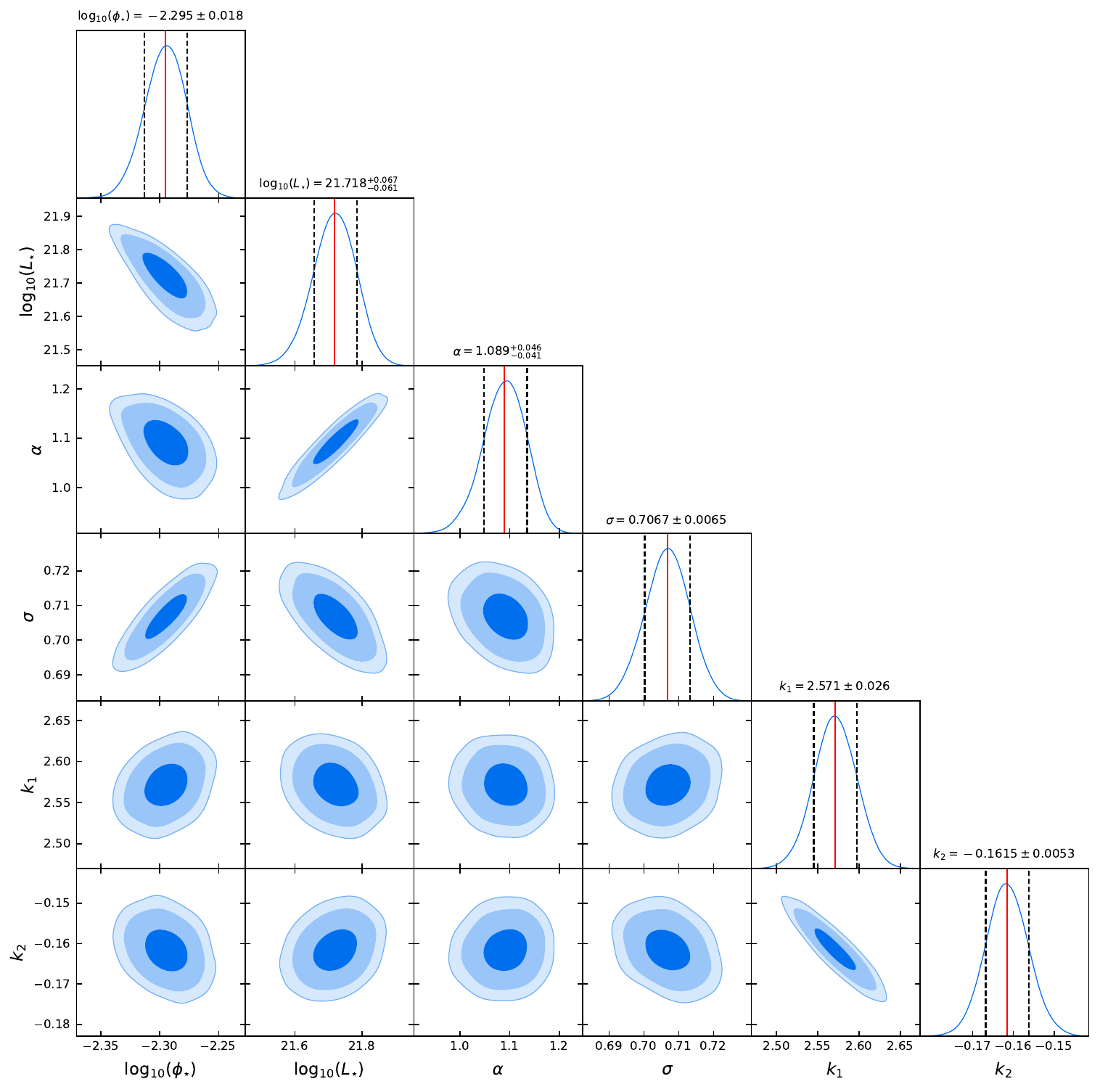}
        \caption{
                ELAIS-N1 field corner plot illustrating the one- and two-dimensional projections of the posterior probability distributions for Model A, derived from the MCMC sampling. The diagonal panels display the marginalized posterior distributions for each parameter, with the 16th and 84th percentiles indicated by vertical dashed lines. The off-diagonal panels present the two-dimensional joint posterior distributions for each parameter pair, with 1$\sigma$, 2$\sigma$, and 3$\sigma$ confidence contours shown as black solid lines. The red vertical solid lines indicate the best-fitting parameter values.}
        \label{fig:en1cornerplotA}
\end{figure*}

\begin{figure*}
        \centering
        \includegraphics[width=\textwidth]{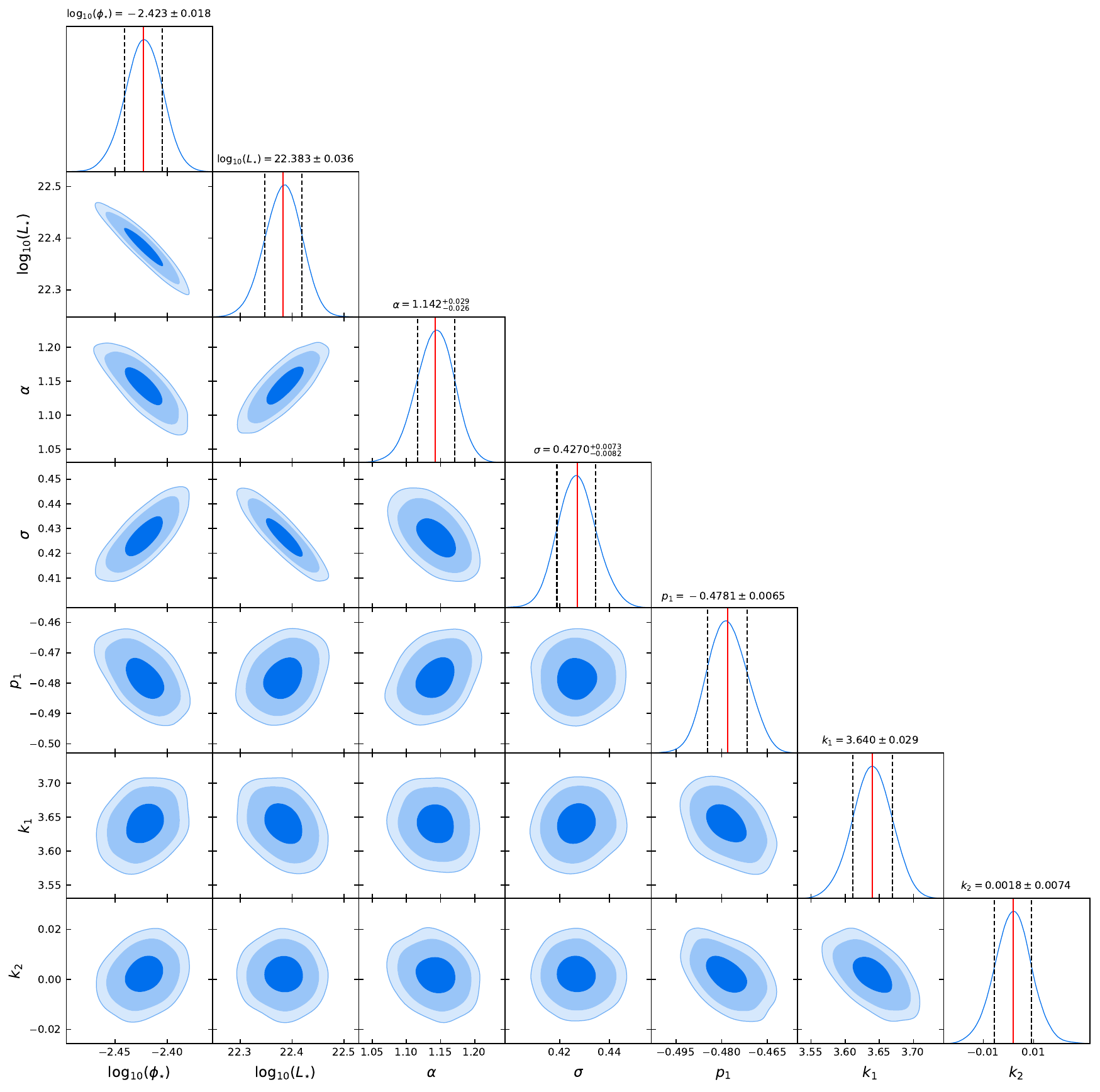}
        \caption{
                Similar to Figure \ref{fig:en1cornerplotA}, but for ELAIS-N1 field Model B.}
        \label{fig:en1cornerplotB}
\end{figure*}

\begin{figure*}
        \centering
        \includegraphics[width=\textwidth]{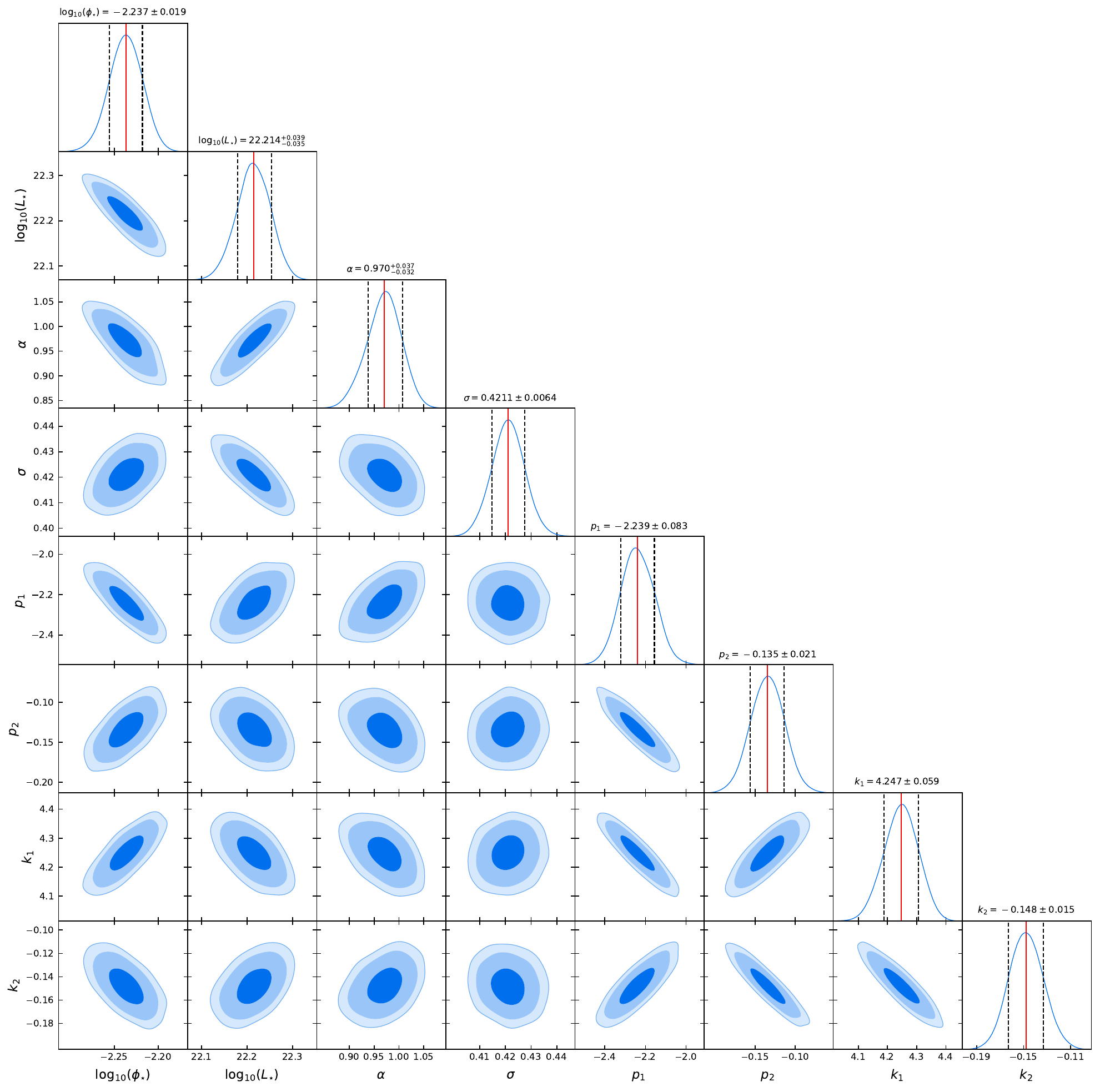}
        \caption{
                Similar to Figure \ref{fig:en1cornerplotA}, but for ELAIS-N1 field Model C.}
        \label{fig:en1cornerplotC}
\end{figure*}

\begin{figure*}
        \centering
        \includegraphics[width=\textwidth]{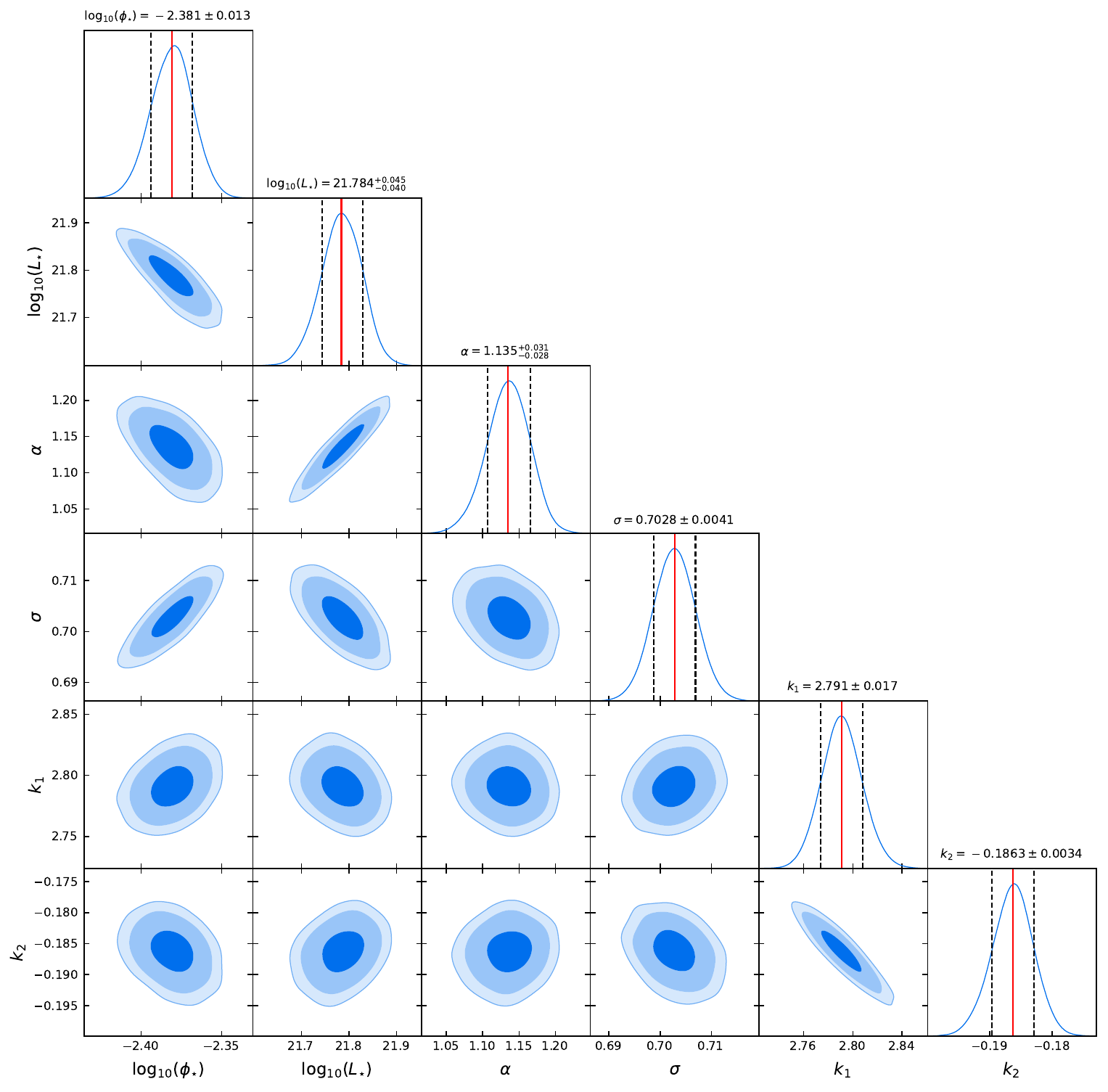}
        \caption{
                Similar to Figure \ref{fig:en1cornerplotA}, but for All fields Model A.}
        \label{fig:allcornerplotA}
\end{figure*}

\begin{figure*}
        \centering
        \includegraphics[width=\textwidth]{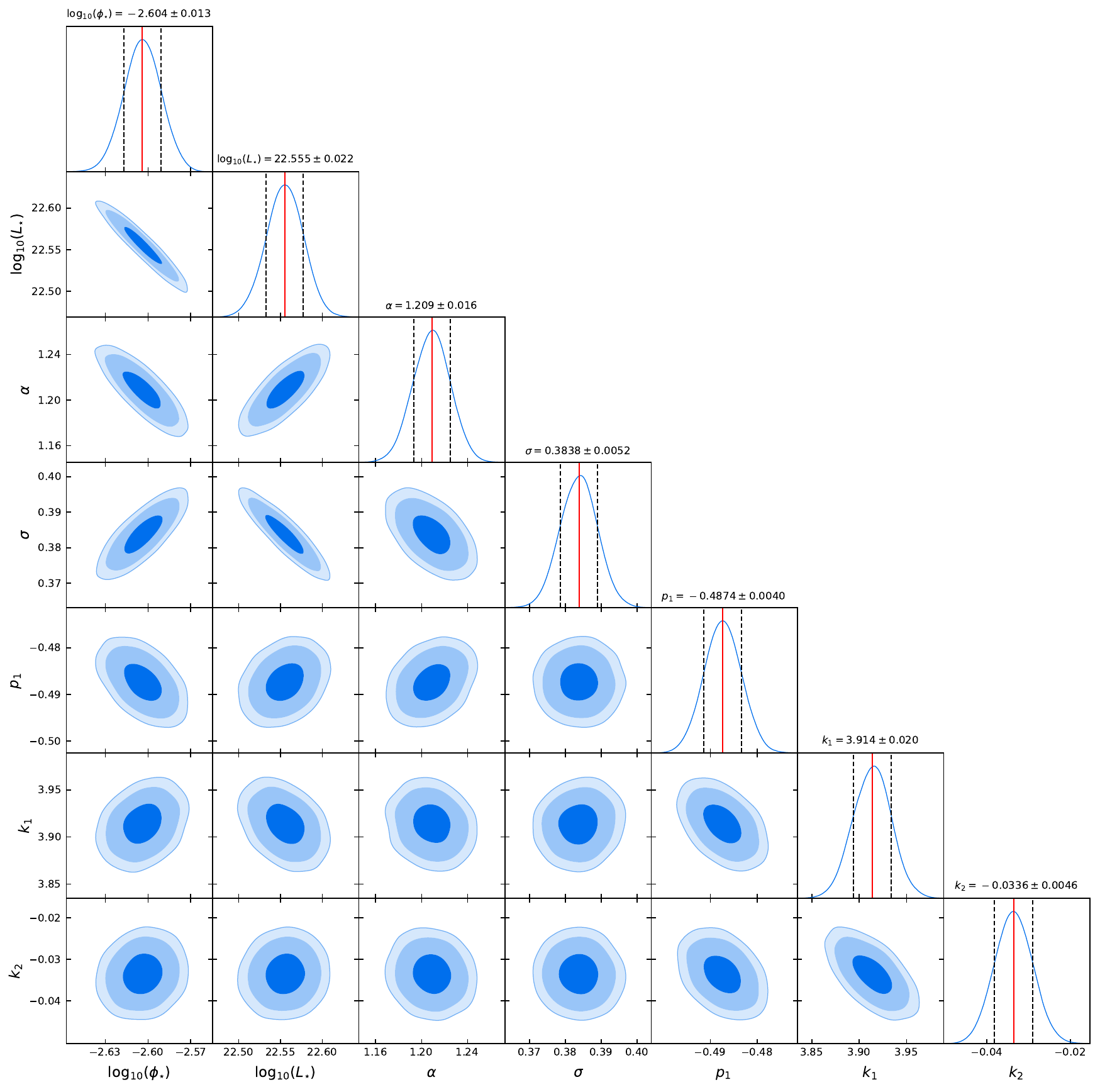}
        \caption{
                Similar to Figure \ref{fig:en1cornerplotA}, but for All fields Model B.}
        \label{fig:allcornerplotB}
\end{figure*}

\begin{figure*}
        \centering
        \includegraphics[width=\textwidth]{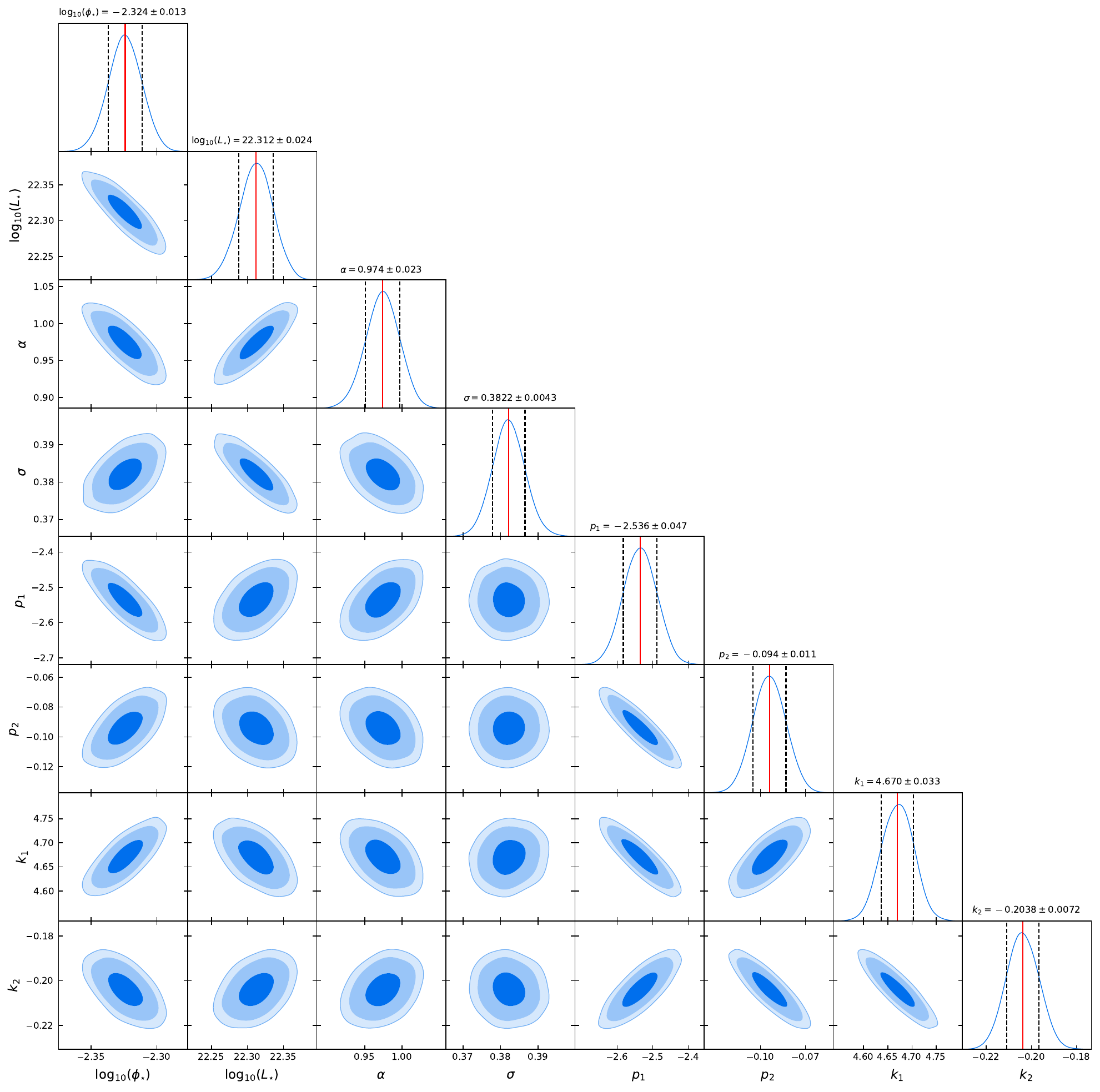}
        \caption{
                Similar to Figure \ref{fig:en1cornerplotA}, but for All fields Model C.}
        \label{fig:allcornerplotC}
\end{figure*}


\begin{thebibliography}{}
\expandafter\ifx\csname natexlab\endcsname\relax\def\natexlab#1{#1}\fi
\providecommand{\url}[1]{\href{#1}{#1}}
\providecommand{\dodoi}[1]{doi:~\href{http://doi.org/#1}{\nolinkurl{#1}}}
\providecommand{\doeprint}[1]{\href{http://ascl.net/#1}{\nolinkurl{http://ascl.net/#1}}}
\providecommand{\doarXiv}[1]{\href{https://arxiv.org/abs/#1}{\nolinkurl{https://arxiv.org/abs/#1}}}

\bibitem[{{Aird} {et~al.}(2017){Aird}, {Coil}, \& {Georgakakis}}]{Aird-2017}
{Aird}, J., {Coil}, A.~L., \& {Georgakakis}, A. 2017, \mnras, 465, 3390,
  \dodoi{10.1093/mnras/stw2932}

\bibitem[{{Aird} {et~al.}(2010){Aird}, {Nandra}, {Laird}, {Georgakakis},
  {Ashby}, {Barmby}, {Coil}, {Huang}, {Koekemoer}, {Steidel}, \&
  {Willmer}}]{2010MNRAS.401.2531A}
{Aird}, J., {Nandra}, K., {Laird}, E.~S., {et~al.} 2010, \mnras, 401, 2531,
  \dodoi{10.1111/j.1365-2966.2009.15829.x}

\bibitem[{{Akaike}(1974)}]{1974ITAC...19..716A}
{Akaike}, H. 1974, IEEE Transactions on Automatic Control, 19, 716

\bibitem[{{Algera} {et~al.}(2020){Algera}, {van der Vlugt}, {Hodge}, {Smail},
  {Novak}, {Radcliffe}, {Riechers}, {R{\"o}ttgering}, {Smol{\v{c}}i{\'c}}, \&
  {Walter}}]{Algera_2020}
{Algera}, H.~S.~B., {van der Vlugt}, D., {Hodge}, J.~A., {et~al.} 2020, \apj,
  903, 139, \dodoi{10.3847/1538-4357/abb77a}

\bibitem[{Ashby {et~al.}(2009)Ashby, Stern, Brodwin, Griffith, Eisenhardt,
  Koz{\l}owski, Kochanek, Bock, Borys, Brand, Brown, Cool, Cooray, Croft, Dey,
  Eisenstein, Gonzalez, Gorjian, Grogin, Ivison, Jacob, Jannuzi, Mainzer,
  Moustakas, R{\"{o}}ttgering, Seymour, Smith, Stanford, Stauffer, Sullivan,
  {Van Breugel}, Willner, \& Wright}]{Ashby2009}
Ashby, M.~L., Stern, D., Brodwin, M., {et~al.} 2009, ApJ, 701, 428,
  \dodoi{10.1088/0004-637X/701/1/428}

\bibitem[{{Bell}(2003)}]{Bell_2003}
{Bell}, E.~F. 2003, \apj, 586, 794, \dodoi{10.1086/367829}

\bibitem[{Best {et~al.}(2023)Best, Kondapally, Williams, Smith, Cochrane, \&
  Duncan}]{Best2023}
Best, P.~N., Kondapally, R., Williams, W., {et~al.} 2023, MNRAS,
  \dodoi{10.1051/0004-6361/202038828}

\bibitem[{{Bouwens} {et~al.}(2015){Bouwens}, {Illingworth}, {Oesch}, {Trenti},
  {Labb{\'e}}, {Bradley}, {Carollo}, {van Dokkum}, {Gonzalez}, {Holwerda},
  {Franx}, {Spitler}, {Smit}, \& {Magee}}]{2015ApJ...803...34B}
{Bouwens}, R.~J., {Illingworth}, G.~D., {Oesch}, P.~A., {et~al.} 2015, \apj,
  803, 34, \dodoi{10.1088/0004-637X/803/1/34}

\bibitem[{{Bouwens} {et~al.}(2021){Bouwens}, {Oesch}, {Stefanon},
  {Illingworth}, {Labb{\'e}}, {Reddy}, {Atek}, {Montes}, {Naidu},
  {Nanayakkara}, {Nelson}, \& {Wilkins}}]{Bouwens_2021}
{Bouwens}, R.~J., {Oesch}, P.~A., {Stefanon}, M., {et~al.} 2021, \aj, 162, 47,
  \dodoi{10.3847/1538-3881/abf83e}

\bibitem[{{Calistro Rivera} {et~al.}(2017){Calistro Rivera}, {Williams},
  {Hardcastle}, {Duncan}, {R{\"o}ttgering}, {Best}, {Br{\"u}ggen}, {Chy{\.z}y},
  {Conselice}, {de Gasperin}, {Engels}, {G{\"u}rkan}, {Intema}, {Jarvis},
  {Mahony}, {Miley}, {Morabito}, {Prandoni}, {Sabater}, {Smith}, {Tasse}, {van
  der Werf}, \& {White}}]{Calistro_2017}
{Calistro Rivera}, G., {Williams}, W.~L., {Hardcastle}, M.~J., {et~al.} 2017,
  \mnras, 469, 3468, \dodoi{10.1093/mnras/stx1040}

\bibitem[{{Chambers} {et~al.}(2016){Chambers}, {Magnier}, {Metcalfe},
  {Flewelling}, {Huber}, {Waters}, {Denneau}, {Draper}, {Farrow}, {Finkbeiner},
  {Holmberg}, {Koppenhoefer}, {Price}, {Rest}, {Saglia}, {Schlafly}, {Smartt},
  {Sweeney}, {Wainscoat}, {Burgett}, {Chastel}, {Grav}, {Heasley}, {Hodapp},
  {Jedicke}, {Kaiser}, {Kudritzki}, {Luppino}, {Lupton}, {Monet}, {Morgan},
  {Onaka}, {Shiao}, {Stubbs}, {Tonry}, {White}, {Ba{\~n}ados}, {Bell},
  {Bender}, {Bernard}, {Boegner}, {Boffi}, {Botticella}, {Calamida},
  {Casertano}, {Chen}, {Chen}, {Cole}, {Deacon}, {Frenk}, {Fitzsimmons},
  {Gezari}, {Gibbs}, {Goessl}, {Goggia}, {Gourgue}, {Goldman}, {Grant},
  {Grebel}, {Hambly}, {Hasinger}, {Heavens}, {Heckman}, {Henderson}, {Henning},
  {Holman}, {Hopp}, {Ip}, {Isani}, {Jackson}, {Keyes}, {Koekemoer}, {Kotak},
  {Le}, {Liska}, {Long}, {Lucey}, {Liu}, {Martin}, {Masci}, {McLean}, {Mindel},
  {Misra}, {Morganson}, {Murphy}, {Obaika}, {Narayan}, {Nieto-Santisteban},
  {Norberg}, {Peacock}, {Pier}, {Postman}, {Primak}, {Rae}, {Rai}, {Riess},
  {Riffeser}, {Rix}, {R{\"o}ser}, {Russel}, {Rutz}, {Schilbach}, {Schultz},
  {Scolnic}, {Strolger}, {Szalay}, {Seitz}, {Small}, {Smith}, {Soderblom},
  {Taylor}, {Thomson}, {Taylor}, {Thakar}, {Thiel}, {Thilker}, {Unger},
  {Urata}, {Valenti}, {Wagner}, {Walder}, {Walter}, {Watters}, {Werner},
  {Wood-Vasey}, \& {Wyse}}]{Chambers2016}
{Chambers}, K.~C., {Magnier}, E.~A., {Metcalfe}, N., {et~al.} 2016, arXiv
  e-prints, arXiv:1612.05560.
\newblock \doarXiv{1612.05560}

\bibitem[{Chen(2017)}]{chen2017tutorial}
Chen, Y.-C. 2017, Biostatistics \& Epidemiology, 1, 161

\bibitem[{{Cochrane} {et~al.}(2023){Cochrane}, {Kondapally}, {Best}, {Sabater},
  {Duncan}, {Smith}, {Hardcastle}, {R{\"o}ttgering}, {Prandoni}, {Haskell},
  {G{\"u}rkan}, \& {Miley}}]{2023MNRAS.523.6082C}
{Cochrane}, R.~K., {Kondapally}, R., {Best}, P.~N., {et~al.} 2023, \mnras, 523,
  6082, \dodoi{10.1093/mnras/stad1602}

\bibitem[{{Condon} {et~al.}(1991){Condon}, {Anderson}, \&
  {Helou}}]{Condon-1991}
{Condon}, J.~J., {Anderson}, M.~L., \& {Helou}, G. 1991, \apj, 376, 95,
  \dodoi{10.1086/170258}

\bibitem[{Davies \& Baddeley(2018)}]{davies2018fast}
Davies, T.~M., \& Baddeley, A. 2018, Statistics and Computing, 28, 937

\bibitem[{{Delvecchio} {et~al.}(2017){Delvecchio}, {Smol{\v c}i{\'c}},
  {Zamorani}, {Lagos}, {Berta}, {Delhaize}, {Baran}, {Alexander}, {Rosario},
  {Gonzalez-Perez}, {Ilbert}, {Lacey}, {Le F{\`e}vre}, {Miettinen}, {Aravena},
  {Bondi}, {Carilli}, {Ciliegi}, {Mooley}, {Novak}, {Schinnerer}, {Capak},
  {Civano}, {Fanidakis}, {Herrera Ruiz}, {Karim}, {Laigle}, {Marchesi},
  {McCracken}, {Middleberg}, {Salvato}, \& {Tasca}}]{Delvecchio_2017}
{Delvecchio}, I., {Smol{\v c}i{\'c}}, V., {Zamorani}, G., {et~al.} 2017, \aap,
  602, A3, \dodoi{10.1051/0004-6361/201629367}

\bibitem[{{Dicken} {et~al.}(2012){Dicken}, {Tadhunter}, {Axon}, {Morganti},
  {Robinson}, {Kouwenhoven}, {Spoon}, {Kharb}, {Inskip}, {Holt}, {Ramos
  Almeida}, \& {Nesvadba}}]{2012ApJ...745..172D}
{Dicken}, D., {Tadhunter}, C., {Axon}, D., {et~al.} 2012, \apj, 745, 172,
  \dodoi{10.1088/0004-637X/745/2/172}

\bibitem[{{Drake} {et~al.}(2013){Drake}, {Simpson}, {Collins}, {James},
  {Baldry}, {Ouchi}, {Jarvis}, {Bonfield}, {Ono}, {Best}, {Dalton}, {Dunlop},
  {McLure}, \& {Smith}}]{Drake-2013}
{Drake}, A.~B., {Simpson}, C., {Collins}, C.~A., {et~al.} 2013, \mnras, 433,
  796, \dodoi{10.1093/mnras/stt775}

\bibitem[{Duncan {et~al.}(2021)Duncan, Kondapally, Brown, Bonato, Best,
  R{\"{o}}ttgering, Bondi, Bowler, Cochrane, G{\"{u}}rkan, Hardcastle, Jarvis,
  Kunert-Bajraszewska, Leslie, Malek, Morabito, O'Sullivan, Prandoni, Sabater,
  Shimwell, Smith, Wang, \& Wolowska}]{Duncan2021}
Duncan, K.~J., Kondapally, R., Brown, M.~J., {et~al.} 2021, A\&A, 648, A4,
  \dodoi{10.1051/0004-6361/202038809}

\bibitem[{{Enia} {et~al.}(2022){Enia}, {Talia}, {Pozzi}, {Cimatti},
  {Delvecchio}, {Zamorani}, {D'Amato}, {Bisigello}, {Gruppioni}, {Rodighiero},
  {Calura}, {Dallacasa}, {Giulietti}, {Barchiesi}, {Behiri}, \&
  {Romano}}]{2022ApJ...927..204E}
{Enia}, A., {Talia}, M., {Pozzi}, F., {et~al.} 2022, \apj, 927, 204,
  \dodoi{10.3847/1538-4357/ac51ca}

\bibitem[{Fan {et~al.}(2001)Fan, Strauss, Schneider, Gunn, Lupton, Becker,
  Davis, Newman, Richards, White, {et~al.}}]{fan2001high}
Fan, X., Strauss, M.~A., Schneider, D.~P., {et~al.} 2001, The Astronomical
  Journal, 121, 54

\bibitem[{Foreman-Mackey {et~al.}(2013)Foreman-Mackey, Hogg, Lang, \&
  Goodman}]{foreman2013emcee}
Foreman-Mackey, D., Hogg, D.~W., Lang, D., \& Goodman, J. 2013, Publications of
  the Astronomical Society of the Pacific, 125, 306

\bibitem[{Gonzalez(2010)}]{Gonzalez2010}
Gonzalez, A.~H. 2010, American Astronomical Society Meeting Abstracts 216

\bibitem[{{Gruppioni} {et~al.}(2013){Gruppioni}, {Pozzi}, {Rodighiero},
  {Delvecchio}, {Berta}, {Pozzetti}, {Zamorani}, {Andreani}, {Cimatti},
  {Ilbert}, {Le Floc'h}, {Lutz}, {Magnelli}, {Marchetti}, {Monaco}, {Nordon},
  {Oliver}, {Popesso}, {Riguccini}, {Roseboom}, {Rosario}, {Sargent},
  {Vaccari}, {Altieri}, {Aussel}, {Bongiovanni}, {Cepa}, {Daddi},
  {Dom{\'\i}nguez-S{\'a}nchez}, {Elbaz}, {F{\"o}rster Schreiber}, {Genzel},
  {Iribarrem}, {Magliocchetti}, {Maiolino}, {Poglitsch}, {P{\'e}rez
  Garc{\'\i}a}, {Sanchez-Portal}, {Sturm}, {Tacconi}, {Valtchanov}, {Amblard},
  {Arumugam}, {Bethermin}, {Bock}, {Boselli}, {Buat}, {Burgarella},
  {Castro-Rodr{\'\i}guez}, {Cava}, {Chanial}, {Clements}, {Conley}, {Cooray},
  {Dowell}, {Dwek}, {Eales}, {Franceschini}, {Glenn}, {Griffin},
  {Hatziminaoglou}, {Ibar}, {Isaak}, {Ivison}, {Lagache}, {Levenson}, {Lu},
  {Madden}, {Maffei}, {Mainetti}, {Nguyen}, {O'Halloran}, {Page}, {Panuzzo},
  {Papageorgiou}, {Pearson}, {P{\'e}rez-Fournon}, {Pohlen}, {Rigopoulou},
  {Rowan-Robinson}, {Schulz}, {Scott}, {Seymour}, {Shupe}, {Smith}, {Stevens},
  {Symeonidis}, {Trichas}, {Tugwell}, {Vigroux}, {Wang}, {Wright}, {Xu},
  {Zemcov}, {Bardelli}, {Carollo}, {Contini}, {Le F{\'e}vre}, {Lilly},
  {Mainieri}, {Renzini}, {Scodeggio}, \& {Zucca}}]{Gruppioni_2013}
{Gruppioni}, C., {Pozzi}, F., {Rodighiero}, G., {et~al.} 2013, \mnras, 432, 23,
  \dodoi{10.1093/mnras/stt308}

\bibitem[{{Gruppioni} {et~al.}(2020){Gruppioni}, {B{\'e}thermin}, {Loiacono},
  {Le F{\`e}vre}, {Capak}, {Cassata}, {Faisst}, {Schaerer}, {Silverman}, {Yan},
  {Bardelli}, {Boquien}, {Carraro}, {Cimatti}, {Dessauges-Zavadsky}, {Ginolfi},
  {Fujimoto}, {Hathi}, {Jones}, {Khusanova}, {Koekemoer}, {Lagache}, {Lemaux},
  {Oesch}, {Pozzi}, {Riechers}, {Rodighiero}, {Romano}, {Talia}, {Vallini},
  {Vergani}, {Zamorani}, \& {Zucca}}]{2020A&A...643A...8G}
{Gruppioni}, C., {B{\'e}thermin}, M., {Loiacono}, F., {et~al.} 2020, \aap, 643,
  A8, \dodoi{10.1051/0004-6361/202038487}

\bibitem[{{G{\"u}rkan} {et~al.}(2018){G{\"u}rkan}, {Hardcastle}, {Smith},
  {Best}, {Bourne}, {Calistro-Rivera}, {Heald}, {Jarvis}, {Prandoni},
  {R{\"o}ttgering}, {Sabater}, {Shimwell}, {Tasse}, \&
  {Williams}}]{2018MNRAS.475.3010G}
{G{\"u}rkan}, G., {Hardcastle}, M.~J., {Smith}, D.~J.~B., {et~al.} 2018,
  \mnras, 475, 3010, \dodoi{10.1093/mnras/sty016}

\bibitem[{{Haarsma} {et~al.}(2000){Haarsma}, {Partridge}, {Windhorst}, \&
  {Richards}}]{Haarsma_2000}
{Haarsma}, D.~B., {Partridge}, R.~B., {Windhorst}, R.~A., \& {Richards}, E.~A.
  2000, \apj, 544, 641, \dodoi{10.1086/317225}

\bibitem[{Hogg(1999)}]{hogg1999distance}
Hogg, D.~W. 1999, arXiv preprint astro-ph/9905116

\bibitem[{{Ishigaki} {et~al.}(2018){Ishigaki}, {Kawamata}, {Ouchi}, {Oguri},
  {Shimasaku}, \& {Ono}}]{2018ApJ...854...73I}
{Ishigaki}, M., {Kawamata}, R., {Ouchi}, M., {et~al.} 2018, \apj, 854, 73,
  \dodoi{10.3847/1538-4357/aaa544}

\bibitem[{Kennicutt~Jr(1998)}]{kennicutt1998star}
Kennicutt~Jr, R.~C. 1998, Annual Review of Astronomy and Astrophysics, 36, 189

\bibitem[{Kondapally {et~al.}(2021)Kondapally, Best, Hardcastle, Nisbet,
  Bonato, Sabater, Duncan, McCheyne, Cochrane, Bowler, Williams, Shimwell,
  Tasse, Croston, Goyal, Jamrozy, Jarvis, Mahatma, R{\"{o}}ttgering, Smith,
  Wo{\l}owska, Bondi, Brienza, Brown, Br{\"{u}}ggen, Chambers, Garrett,
  G{\"{u}}rkan, Huber, Kunert-Bajraszewska, Magnier, Mingo, Mostert,
  Nikiel-Wroczy{\'{n}}ski, O'Sullivan, Paladino, Ploeckinger, Prandoni,
  Rosenthal, Schwarz, Shulevski, Wagenveld, \& Wang}]{Kondapally2021}
Kondapally, R., Best, P.~N., Hardcastle, M.~J., {et~al.} 2021, A\&A, 648, A3,
  \dodoi{10.1051/0004-6361/202038813}

\bibitem[{Lawrence {et~al.}(2007)Lawrence, Warren, Almaini, Edge, Hambly,
  Jameson, Lucas, Casali, Adamson, Dye, Emerson, Foucaud, Hewett, Hirst,
  Hodgkin, Irwin, Lodieu, McMahon, Simpson, Smail, Mortlock, \&
  Folger}]{Lawrence2007}
Lawrence, A., Warren, S.~J., Almaini, O., {et~al.} 2007, MNRAS, 379, 1599,
  \dodoi{10.1111/j.1365-2966.2007.12040.x}

\bibitem[{Lonsdale {et~al.}(2003)Lonsdale, Smith, Robinson, Surace, Shupe, Xu,
  Oliver, Padgett, Fang, Conrow, Gautier, Griffin, Hacking, Masci, Morrison,
  Linger, Owen, Fournon, Pierre, Puetter, Stacey, Castro, Del, Polletta,
  Farrah, Jarrett, Publications, Society, August, Lonsdale, Smith,
  Rowan-robinson, Surace, Shupe, Xu, Oliver, Padgett, Fang, Conrow,
  Franceschini, Gautier, Griffin, Hacking, Masci, Morrison, Linger, Owen, Pe,
  Pierre, Puetter, Stacey, Castro, Del, Polletta, Farrah, Jarrett, Frayer,
  Siana, Babbedge, Dye, Fox, Gonzalez-solares, Salaman, Berta, \&
  Condon}]{Lonsdale2003}
Lonsdale, C.~J., Smith, H.~E., Robinson, M.~R., {et~al.} 2003, PASP, 115, 897

\bibitem[{{Madau} \& {Dickinson}(2014)}]{Madau_2014}
{Madau}, P., \& {Dickinson}, M. 2014, \araa, 52, 415,
  \dodoi{10.1146/annurev-astro-081811-125615}

\bibitem[{{Madau} {et~al.}(1996){Madau}, {Ferguson}, {Dickinson}, {Giavalisco},
  {Steidel}, \& {Fruchter}}]{Madau_1996MNRAS.283.1388M}
{Madau}, P., {Ferguson}, H.~C., {Dickinson}, M.~E., {et~al.} 1996, \mnras, 283,
  1388, \dodoi{10.1093/mnras/283.4.1388}

\bibitem[{Magnelli {et~al.}(2015)Magnelli, Ivison, Lutz, Valtchanov, Farrah,
  Berta, Bertoldi, Bock, Cooray, Ibar, {et~al.}}]{magnelli2015far}
Magnelli, B., Ivison, R., Lutz, D., {et~al.} 2015, Astronomy \& Astrophysics,
  573, A45

\bibitem[{{Malefahlo} {et~al.}(2022){Malefahlo}, {Jarvis}, {Santos}, {White},
  {Adams}, \& {Bowler}}]{2022MNRAS.509.4291M}
{Malefahlo}, E.~D., {Jarvis}, M.~J., {Santos}, M.~G., {et~al.} 2022, \mnras,
  509, 4291, \dodoi{10.1093/mnras/stab3242}

\bibitem[{Marshall {et~al.}(1983)Marshall, Tananbaum, Avni, \&
  Zamorani}]{marshall1983analysis}
Marshall, H., Tananbaum, H., Avni, Y., \& Zamorani, G. 1983, The Astrophysical
  Journal, 269, 35

\bibitem[{Martin {et~al.}(2005)Martin, Fanson, Schiminovich, Morrissey,
  Friedman, Barlow, Conrow, Grange, Jelinsky, Milliard, Siegmund, Bianchi,
  Byun, Donas, Forster, Heckman, Lee, Madore, Malina, Neff, Rich, Small,
  Surber, Szalay, Welsh, Wyder, \& Al}]{Martin2005}
Martin, D.~C., Fanson, J., Schiminovich, D., {et~al.} 2005, ApJ, 619, L1

\bibitem[{{McLeod} {et~al.}(2016){McLeod}, {McLure}, \&
  {Dunlop}}]{2016MNRAS.459.3812M}
{McLeod}, D.~J., {McLure}, R.~J., \& {Dunlop}, J.~S. 2016, \mnras, 459, 3812,
  \dodoi{10.1093/mnras/stw904}

\bibitem[{{McLure} {et~al.}(2013){McLure}, {Dunlop}, {Bowler}, {Curtis-Lake},
  {Schenker}, {Ellis}, {Robertson}, {Koekemoer}, {Rogers}, {Ono}, {Ouchi},
  {Charlot}, {Wild}, {Stark}, {Furlanetto}, {Cirasuolo}, \&
  {Targett}}]{Mclure_2013}
{McLure}, R.~J., {Dunlop}, J.~S., {Bowler}, R.~A.~A., {et~al.} 2013, \mnras,
  432, 2696, \dodoi{10.1093/mnras/stt627}

\bibitem[{{Mingo} {et~al.}(2014){Mingo}, {Hardcastle}, {Croston}, {Dicken},
  {Evans}, {Morganti}, \& {Tadhunter}}]{2014MNRAS.440..269M}
{Mingo}, B., {Hardcastle}, M.~J., {Croston}, J.~H., {et~al.} 2014, \mnras, 440,
  269, \dodoi{10.1093/mnras/stu263}

\bibitem[{Mohan \& Rafferty(2015)}]{Mohan2015}
Mohan, N., \& Rafferty, D. 2015, Astrophysics Source Code Library.
\newblock \url{ascl:1502.007}

\bibitem[{{Novak} {et~al.}(2017){Novak}, {Smol{\v{c}}i{\'c}}, {Delhaize},
  {Delvecchio}, {Zamorani}, {Baran}, {Bondi}, {Capak}, {Carilli}, {Ciliegi},
  {Civano}, {Ilbert}, {Karim}, {Laigle}, {Le F{\`e}vre}, {Marchesi},
  {McCracken}, {Miettinen}, {Salvato}, {Sargent}, {Schinnerer}, \&
  {Tasca}}]{2017A&A...602A...5N}
{Novak}, M., {Smol{\v{c}}i{\'c}}, V., {Delhaize}, J., {et~al.} 2017, \aap, 602,
  A5, \dodoi{10.1051/0004-6361/201629436}

\bibitem[{Ocran {et~al.}(2020)Ocran, Taylor, Vaccari, Ishwara-Chandra,
  Prandoni, Prescott, \& Mancuso}]{ocran2020cosmic}
Ocran, E., Taylor, A., Vaccari, M., {et~al.} 2020, Monthly Notices of the Royal
  Astronomical Society, 491, 5911

\bibitem[{{Ocran} {et~al.}(2020){Ocran}, {Taylor}, {Vaccari},
  {Ishwara-Chandra}, {Prandoni}, {Prescott}, \&
  {Mancuso}}]{2020MNRAS.491.5911O}
{Ocran}, E.~F., {Taylor}, A.~R., {Vaccari}, M., {et~al.} 2020, \mnras, 491,
  5911, \dodoi{10.1093/mnras/stz3401}

\bibitem[{{Oesch} {et~al.}(2018){Oesch}, {Bouwens}, {Illingworth}, {Labb{\'e}},
  \& {Stefanon}}]{Oesch_2018}
{Oesch}, P.~A., {Bouwens}, R.~J., {Illingworth}, G.~D., {Labb{\'e}}, I., \&
  {Stefanon}, M. 2018, \apj, 855, 105, \dodoi{10.3847/1538-4357/aab03f}

\bibitem[{Oliver {et~al.}(2012)Oliver, Bock, Altieri, Amblard, Arumugam,
  Aussel, Babbedge, Beelen, B{\'{e}}thermin, Blain, Boselli, Bridge, Brisbin,
  Buat, Burgarella, Castro-Rodr{\'{i}}guez, Cava, Chanial, Cirasuolo, Clements,
  Conley, Conversi, Cooray, Dowell, Dubois, Dwek, Dye, Eales, Elbaz, Farrah,
  Feltre, Ferrero, Fiolet, Fox, Franceschini, Gear, Giovannoli, Glenn, Gong,
  {Gonz{\'{a}}lez Solares}, Griffin, Halpern, Harwit, Hatziminaoglou, Heinis,
  Hurley, Hwang, Hyde, Ibar, Ilbert, Isaak, Ivison, Lagache, {Le Floc'h},
  Levenson, Faro, Lu, Madden, Maffei, Magdis, Mainetti, Marchetti, Marsden,
  Marshall, Mortier, Nguyen, O'Halloran, Omont, Page, Panuzzo, Papageorgiou,
  Patel, Pearson, P{\'{e}}rez-Fournon, Pohlen, Rawlings, Raymond, Rigopoulou,
  Riguccini, Rizzo, Rodighiero, Roseboom, Rowan-Robinson, {S{\'{a}}nchez
  Portal}, Schulz, Scott, Seymour, Shupe, Smith, Stevens, Symeonidis, Trichas,
  Tugwell, Vaccari, Valtchanov, Vieira, Viero, Vigroux, Wang, Ward, Wardlow,
  Wright, Xu, \& Zemcov}]{Oliver2012}
Oliver, S.~J., Bock, J., Altieri, B., {et~al.} 2012, MNRAS, 424, 1614,
  \dodoi{10.1111/j.1365-2966.2012.20912.x}

\bibitem[{{Ouchi} {et~al.}(2010){Ouchi}, {Shimasaku}, {Furusawa}, {Saito},
  {Yoshida}, {Akiyama}, {Ono}, {Yamada}, {Ota}, {Kashikawa}, {Iye}, {Kodama},
  {Okamura}, {Simpson}, \& {Yoshida}}]{Ouchi-2010}
{Ouchi}, M., {Shimasaku}, K., {Furusawa}, H., {et~al.} 2010, \apj, 723, 869,
  \dodoi{10.1088/0004-637X/723/1/869}

\bibitem[{{Padovani}(2016)}]{Padovani_2016}
{Padovani}, P. 2016, Astronomy and Astrophysics Review, 24, 13,
  \dodoi{10.1007/s00159-016-0098-6}

\bibitem[{{Pritchard} \& {Loeb}(2010)}]{2010Natur.468..772P}
{Pritchard}, J., \& {Loeb}, A. 2010, \nat, 468, 772, \dodoi{10.1038/468772b}

\bibitem[{{Riechers} {et~al.}(2013){Riechers}, {Bradford}, {Clements},
  {Dowell}, {P{\'e}rez-Fournon}, {Ivison}, {Bridge}, {Conley}, {Fu}, {Vieira},
  {Wardlow}, {Calanog}, {Cooray}, {Hurley}, {Neri}, {Kamenetzky}, {Aguirre},
  {Altieri}, {Arumugam}, {Benford}, {B{\'e}thermin}, {Bock}, {Burgarella},
  {Cabrera-Lavers}, {Chapman}, {Cox}, {Dunlop}, {Earle}, {Farrah}, {Ferrero},
  {Franceschini}, {Gavazzi}, {Glenn}, {Solares}, {Gurwell}, {Halpern},
  {Hatziminaoglou}, {Hyde}, {Ibar}, {Kov{\'a}cs}, {Krips}, {Lupu}, {Maloney},
  {Martinez-Navajas}, {Matsuhara}, {Murphy}, {Naylor}, {Nguyen}, {Oliver},
  {Omont}, {Page}, {Petitpas}, {Rangwala}, {Roseboom}, {Scott}, {Smith},
  {Staguhn}, {Streblyanska}, {Thomson}, {Valtchanov}, {Viero}, {Wang},
  {Zemcov}, \& {Zmuidzinas}}]{Riechers_2013}
{Riechers}, D.~A., {Bradford}, C.~M., {Clements}, D.~L., {et~al.} 2013, \nat,
  496, 329, \dodoi{10.1038/nature12050}

\bibitem[{{Rowan-Robinson} {et~al.}(2016){Rowan-Robinson}, {Oliver}, {Wang},
  {Farrah}, {Clements}, {Gruppioni}, {Marchetti}, {Rigopoulou}, \&
  {Vaccari}}]{RowanRobinson_2016}
{Rowan-Robinson}, M., {Oliver}, S., {Wang}, L., {et~al.} 2016, \mnras, 461,
  1100, \dodoi{10.1093/mnras/stw1169}

\bibitem[{Sabater {et~al.}(2021)Sabater, Best, Tasse, Hardcastle, Shimwell,
  Nisbet, Jelic, Callingham, R{\"{o}}ttgering, Bonato, Bondi, Ciardi, Cochrane,
  Jarvis, Kondapally, Koopmans, O'Sullivan, Prandoni, Schwarz, Smith, Wang,
  Williams, \& Zaroubi}]{Sabater2021}
Sabater, J., Best, P.~N., Tasse, C., {et~al.} 2021, A\&A, 648, A2,
  \dodoi{10.1051/0004-6361/202038828}

\bibitem[{{Sadler} {et~al.}(1989){Sadler}, {Jenkins}, \&
  {Kotanyi}}]{Sadler_1989}
{Sadler}, E.~M., {Jenkins}, C.~R., \& {Kotanyi}, C.~G. 1989, \mnras, 240, 591,
  \dodoi{10.1093/mnras/240.3.591}

\bibitem[{Saunders {et~al.}(1990)Saunders, Rowan-Robinson, Lawrence,
  Efstathiou, Kaiser, Ellis, \& Frenk}]{saunders199060}
Saunders, W., Rowan-Robinson, M., Lawrence, A., {et~al.} 1990, Monthly Notices
  of the Royal Astronomical Society, 242, 318

\bibitem[{{Schmidt}(1968)}]{1968ApJ...151..393S}
{Schmidt}, M. 1968, \apj, 151, 393, \dodoi{10.1086/149446}

\bibitem[{{Schober} {et~al.}(2015){Schober}, {Schleicher}, \&
  {Klessen}}]{Schober-2015}
{Schober}, J., {Schleicher}, D.~R.~G., \& {Klessen}, R.~S. 2015, \mnras, 446,
  2, \dodoi{10.1093/mnras/stu1999}

\bibitem[{Schwarz(1978)}]{Schwarz1978}
Schwarz, G. 1978, The Annals of Statistics, 6, 461 ,
  \dodoi{10.1214/aos/1176344136}

\bibitem[{Singal {et~al.}(2014)Singal, Ko, \& Petrosian}]{singal2014gamma}
Singal, J., Ko, A., \& Petrosian, V. 2014, The Astrophysical Journal, 786, 109

\bibitem[{Singal {et~al.}(2013)Singal, Petrosian, Lawrence,
  {et~al.}}]{singal2013radio}
Singal, J., Petrosian, V., Lawrence, A., {et~al.} 2013, The Astrophysical
  Journal, 764, 43

\bibitem[{{Smail} {et~al.}(1997){Smail}, {Ivison}, \& {Blain}}]{Smail_1997}
{Smail}, I., {Ivison}, R.~J., \& {Blain}, A.~W. 1997, \apjl, 490, L5,
  \dodoi{10.1086/311017}

\bibitem[{{Smith} {et~al.}(2021){Smith}, {Haskell}, {G{\"u}rkan}, {Best},
  {Hardcastle}, {Kondapally}, {Williams}, {Duncan}, {Cochrane}, {McCheyne},
  {R{\"o}ttgering}, {Sabater}, {Shimwell}, {Tasse}, {Bonato}, {Bondi},
  {Jarvis}, {Leslie}, {Prandoni}, \& {Wang}}]{2021A&A...648A...6S}
{Smith}, D.~J.~B., {Haskell}, P., {G{\"u}rkan}, G., {et~al.} 2021, \aap, 648,
  A6, \dodoi{10.1051/0004-6361/202039343}

\bibitem[{Smol{\v{c}}i{\'c} {et~al.}(2009)Smol{\v{c}}i{\'c}, Zamorani,
  Schinnerer, Bardelli, Bondi, B{\^\i}rzan, Carilli, Ciliegi, Elvis, Impey,
  {et~al.}}]{smolvcic2009cosmic}
Smol{\v{c}}i{\'c}, V., Zamorani, G., Schinnerer, E., {et~al.} 2009, The
  Astrophysical Journal, 696, 24

\bibitem[{{Smol{\v{c}}i{\'c}} {et~al.}(2009){Smol{\v{c}}i{\'c}}, {Schinnerer},
  {Zamorani}, {Bell}, {Bondi}, {Carilli}, {Ciliegi}, {Mobasher}, {Paglione},
  {Scodeggio}, \& {Scoville}}]{2009ApJ...690..610S}
{Smol{\v{c}}i{\'c}}, V., {Schinnerer}, E., {Zamorani}, G., {et~al.} 2009, \apj,
  690, 610, \dodoi{10.1088/0004-637X/690/1/610}

\bibitem[{{Tabatabaei} {et~al.}(2017){Tabatabaei}, {Schinnerer}, {Krause},
  {Dumas}, {Meidt}, {Damas-Segovia}, {Beck}, {Murphy}, {Mulcahy}, {Groves},
  {Bolatto}, {Dale}, {Galametz}, {Sandstrom}, {Boquien}, {Calzetti},
  {Kennicutt}, {Hunt}, {De Looze}, \& {Pellegrini}}]{Tabatabaei_2017}
{Tabatabaei}, F.~S., {Schinnerer}, E., {Krause}, M., {et~al.} 2017, \apj, 836,
  185, \dodoi{10.3847/1538-4357/836/2/185}

\bibitem[{Tasse {et~al.}(2021)Tasse, Shimwell, Hardcastle, O'Sullivan, van
  Weeren, Best, Bester, Hugo, Smirnov, Sabater, Calistro-Rivera, de~Gasperin,
  Morabito, R{\"{o}}ttgering, Williams, Bonato, Bondi, Botteon, Br{\"{u}}ggen,
  Brunetti, Chy~zy, Garrett, G{\"{u}}rkan, Jarvis, Kondapally, Mandal,
  Prandoni, Repetti, Retana-Montenegro, Schwarz, Shulevski, \&
  Wiaux}]{Tasse2021}
Tasse, C., Shimwell, T., Hardcastle, M.~J., {et~al.} 2021, A\&A, 648, A1,
  \dodoi{10.1051/0004-6361/202038804}

\bibitem[{{van der Vlugt} {et~al.}(2022){van der Vlugt}, {Hodge}, {Algera},
  {Smail}, {Leslie}, {Radcliffe}, {Riechers}, \&
  {R{\"o}ttgering}}]{2022ApJ...941...10V}
{van der Vlugt}, D., {Hodge}, J.~A., {Algera}, H.~S.~B., {et~al.} 2022, \apj,
  941, 10, \dodoi{10.3847/1538-4357/ac99db}

\bibitem[{{Wang} {et~al.}(2024){Wang}, {Yuan}, {Yu}, \&
  {Mao}}]{2024A&A...683A.174W}
{Wang}, W., {Yuan}, Z., {Yu}, H., \& {Mao}, J. 2024, \aap, 683, A174,
  \dodoi{10.1051/0004-6361/202347746}

\bibitem[{Wasserman(2006)}]{wasserman2006all}
Wasserman, L. 2006, All of nonparametric statistics (Springer Science \&
  Business Media)

\bibitem[{{Weymann} {et~al.}(1999){Weymann}, {Storrie-Lombardi}, {Sawicki}, \&
  {Brunner}}]{1999ASPC..191.....W}
{Weymann}, R., {Storrie-Lombardi}, L., {Sawicki}, M., \& {Brunner}, R., eds.
  1999, Astronomical Society of the Pacific Conference Series, Vol. 191,
  {Photometric Redshifts and the Detection of High Redshift Galaxies}

\bibitem[{Willott {et~al.}(2001)Willott, Rawlings, Blundell, Lacy, \&
  Eales}]{willott2001radio}
Willott, C.~J., Rawlings, S., Blundell, K.~M., Lacy, M., \& Eales, S.~A. 2001,
  Monthly Notices of the Royal Astronomical Society, 322, 536

\bibitem[{Yuan {et~al.}(2020)Yuan, Jarvis, \& Wang}]{yuan2020flexible}
Yuan, Z., Jarvis, M.~J., \& Wang, J. 2020, The Astrophysical Journal Supplement
  Series, 248, 1

\bibitem[{Yuan {et~al.}(2016)Yuan, Wang, Zhou, \& Mao}]{yuan2016mixture}
Yuan, Z., Wang, J., Zhou, M., \& Mao, J. 2016, The Astrophysical Journal, 820,
  65

\bibitem[{Yuan {et~al.}(2017)Yuan, Wang, Zhou, Qin, \& Mao}]{yuan2017mixture}
Yuan, Z., Wang, J., Zhou, M., Qin, L., \& Mao, J. 2017, The Astrophysical
  Journal, 846, 78

\bibitem[{Yuan {et~al.}(2022)Yuan, Zhang, Wang, Cheng, \&
  Wang}]{yuan2022flexible}
Yuan, Z., Zhang, X., Wang, J., Cheng, X., \& Wang, W. 2022, The Astrophysical
  Journal Supplement Series, 260, 10

\end{thebibliography}
\end{document}